\documentclass[useAMS,usenatbib, 
]{mnras}
\usepackage{natbib}
\usepackage{graphicx}
\usepackage{epsfig}
\usepackage{amssymb}
\usepackage{amsmath}
\usepackage{natbib}
\usepackage{times}

\usepackage{comment}

\usepackage{bm}

\usepackage[english]{babel}

\title[Gravitational instability of gas-dust mixture
]{Dynamic role of dust in formation of molecular clouds}
\author[V.V. Zhuravlev]{V. V. Zhuravlev$^{1}$\thanks{E-mail:
zhuravlev@sai.msu.ru} \\
$^{1}$Sternberg Astronomical Institute, Lomonosov Moscow State University, Universitetskij pr., 13, Moscow 119234, Russia}

\begin{document}

\date{%Accepted 
%1988 December 15. Received ; in original form 
}

\pagerange{\pageref{firstpage}--\pageref{lastpage}} \pubyear{2017}

\maketitle

\label{firstpage}

\defcitealias{squire-2018_acoustic}{HS18}

\begin{abstract}

Dust is the usual minor component of the interstellar medium. Its dynamic role in the contraction of 
the diffuse gas into molecular clouds is commonly assumed to be negligible because of the small mass fraction, 
$f \simeq 0.01$. 
However, as shown in this study, the collective motion of dust grains with respect to the gas 
may considerably contribute to the destabilisation of the medium
on scales $\lambda \lesssim \lambda_J$, where $\lambda_J$ is the Jeans length-scale.
The linear perturbations of the uniform self-gravitating gas at rest are marginally stable at $\lambda \simeq \lambda_J$, 
but as soon as the drift of grains is taken into account,
they begin growing at a rate approximately equal to $(f \tau)^{1/3} t^{-1}_{ff}$, where $\tau$ is the stopping time of grains expressed in units of the free fall time of the cloud, $t_{ff}$. The physical mechanism responsible for such a weak dependence of the growth rate on $f$ is the resonance of heavy sound waves stopped by the self-gravity of gas with
weak gravitational attraction caused by perturbations of the dust fraction. 
Once there is stationary subsonic bulk drift of the dust, 
the growing gas-dust perturbations at $\lambda < \lambda_J$ become waves propagating with the drift velocity
projected onto the wavevector. Their growth has a resonant nature as well and the growth rate is substantially larger 
than that of the recently discovered resonant instability of gas-dust mixture in the absence of self-gravity.
The new instabilities can facilitate gravitational contraction of cold interstellar gas into clouds and 
additionally produce dusty domains of sub-Jeans size at different stages of molecular cloud formation and evolution. 

\end{abstract}


\section{Introduction}

%Сначала - про симуляции молекулярных облаков. 
%There is evidence that gravitational contraction is important at all stages/scales of the formation and evolution of
%star forming regions, see \citet{vazquez-semadeni-2019}.
%The most recent review on the formation of star clusters \citet{krause-2020}.

There is growing observational and numerical evidence that star forming regions may be in a state of global 
gravitational contraction, see \citet{vazquez-semadeni-2019}. The supersonic collisions of flows of warm diffuse 
atomic gas simulated with both self-gravity and cooling exhibit the hierarchical collapse of the
turbulent medium, as was shown by \citet{vazquez-semadeni-2007} and \citet{vazquez-semadeni-2015} for example.
This implies that gravitational instability (GI hereafter) manifests itself
in a wide range of sufficiently large scales during the evolution of molecular clouds.
%Сюда обзор GI разных глобальных конфигураций.
Theoretical work has revealed that flattened dense structures form as a result of large collisions of diffuse matter. 
Later on, they give birth to filaments which then fragment into multiple cores.
This scenario is provided by the dynamical 
instability of self-gravitating layers, cylinders and spheres, respectively. 
The linear stability analysis of these idealised configurations  
(e.g. \citealt{ledoux-1951, chandrasekhar-1953, bonnor-1956, elmegreen-1978, nagasawa-1987, pudritz-2000} ) as well as the corresponding non-linear solutions 
(e.g. \citealt{larson-1969, penston-1969, inutsuka-2000, miyama-1987, inutsuka-1997} and many others) confirm this view. 
At the same time, as was noted by \citet{larson-1985}, the specific geometry of self-gravitating objects 
is not crucial for the instability condition, which does not differ much from the basic one derived 
for the unbounded uniform medium. 
In the latter case, the study of GI goes back to \citet{jeans-1902}, 
who established that plane-wave perturbations on such a background having finite pressure 
are heavy sound waves propagating at the subsonic
velocity, which vanishes as the wavelength approaches the value now referred to as the Jeans length. Perturbations with 
scale larger than the Jeans length are the growing and damping static waves. Thus, the critical scales for GI 
of realistic configurations mentioned above are always similar to the Jeans scale, which includes 
typical speed of sound and density chosen appropriately for the corresponding configuration. 
However, the most unstable scale for realistic configurations has a finite value in contrast to the Jeans result,
when the largest growth rate (corresponding to the inverse free fall time) manifests at the infinitely large scale. 
The largest growth rates for GI of realistic configurations are commonly the fractions of the 
inverse free fall time.

Dust is a component of the diffuse interstellar medium (ISM hereafter) usually considered as an agent for its 
thermal and chemical 
evolution on the way to star formation \citep{girichidis-2020, krause-2020}.
The measured mass fraction of dust with respect to gas in the Milky Way is around $0.01$ \citep{draine-2011}. 
It might seem that such a small value rules out the possibility that the dust could dynamically affect 
the formation of dense clouds of neutral/molecular hydrogen or even the subsequent collapse of prestellar cores. 

Until recently, dust has been considered as only a passive constituent of the clouds 
which, however, could be only partially coupled to the gas for sufficiently large grains. 
This feature may lead to concentration of dust. Indeed, grains 
dynamically interact with the gas due to the aerodynamic drag \citep{whipple-1972, weidenschilling-1977}
parametrised by the characteristic stopping time, which is the time over which a particular grain loses 
its initial velocity in the absence of other forces.
As the stopping time becomes longer, grains may gain higher velocity relative to the gas. 
First, stationary bulk drift of the grains under the action of the anisotropic interstellar radiation field may occur.
This is produced by the radiation pressure force along with photoelectric and photodesorption forces,
see \citet{weingartner-2001}. They show that sufficiently large grains, up to the micron size, experience 
considerable subsonic drift in the warm and cold ISM. This effect can be enhanced up to the transonic
and even supersonic drift in the vicinity of bright sources such as AGN and starburst regions. 
Next, the dust sinks down to pressure maxima. This feature is widely known in the context of dust dynamics
in protoplanetary discs, as it causes a global inward radial drift and vertical settling of solids along with
their local concentration in axisymmetric pressure bumps/zonal flows or in the long-living vortices 
generated by the turbulence \citep{johansen-2014}. 
%The former process represents (locally) stationary drift of the dust, while the latter is of a transient nature. 
Dust sedimentation in the potential well of an interstellar gas cloud in hydrostatic equilibrium is another example 
of the dust drift considered by \citet{flannery-1978}. It was shown that 
micron-sized grains settle to the centre of a cold uniform cloud at a characteristic time 
not much exceeding the free fall time of the cloud. 
%У них то же самое, что у нас, но без фидбека --- это соответствует линейной росту \delta у нас.
%При этом, характерное время оседания каждой пылинки у них \tau^{-1} от времени свободного падения, что
%при малом f даже меньше, чем наше время развития неустойчивости. Оседание пылинки происходит по экспоненциальному 
%закону, но неустойчивости нет, а за характерное время концентрация пыли возрастает в 2 раза просто.
In the past few years the relative motion of dust in turbulent clouds has been studied employing the numerical simulations,
%found in citations of Larson-1969.
see \citet{hopkins-2016}, \citet{hopkins-2017}, \citet{laibe-2017}, \citet{monceau-baroux-2017} and 
\citet{mattsson-2019}. These studies revealed the significant fluctuations of the dust density of (sub-)micron-sized
grains at sub-parsec scales, though there remains a discrepancy in the magnitude of overdensities obtained by
various numerical methods.
In contrast to similar problem in protoplanetary discs, turbulence in molecular clouds is supersonic, which 
complicates the underlying physics. The dust is dragged by the compressible gas, which implies that the dust clumping 
is additionally produced by the compression of the gas-dust mixture, as well as by fluctuations of the drag itself. 
Note that currently turbulent concentration of interstellar dust has not been not studied in self-gravitating 
configurations. Also, no one has considered how the dust back-reaction on gas affects the concentration of dust 
due to the externally driven turbulence.

At the same time, \citet{squire-2018_gen} recognised a class of resonant dynamical instabilities inherent in
the partially coupled gas-dust mixture when gas and dust interact with each other via aerodynamic drag. 
For resonant instability to operate, the phase speed of some wave existing in the gas should match 
the projection of the drift velocity of dust onto the wavevector, thus, generally dust must flow through the gas. 
The resonant instability can manifest itself in various objects, e.g. in
protoplanetary discs \citep{squire_2018} or hot magnetised circumstellar medium such as stellar coronae and HII regions \citep{squire_2018_magnetic}. As the dust drift becomes nearly sonic or even supersonic, the most basic case of 
the acoustic resonant instability is realised, see \citet{squire-2018_acoustic} 
(\citetalias{squire-2018_acoustic} hereafter), which may be relevant in the neutral circumstellar medium.
For the non-linear outcome of this particular instability see \citet{squire-non-linear-acoustic-2019}.
It is important that the resonant instability of the gas-dust mixture is characterised by a weak dependence 
on the dust fraction. Its growth rate usually scales as the square or even the cube root of the dust fraction. 
At least for the particular model of the dust streaming in protoplanetary disc, this feature was explained by the mode
coupling of gas-dust perturbations, see \citet{zhuravlev-2019}.
This implies that the resonant instability may be important in application to the ISM, 
where the dust fraction is typically small. Furthermore, it may not only provide the dust clumping but also 
significantly affect the gas dynamics. 

%%%%%%%%%%%%%%%%%%%%%%%%%%%%%%%%%%

This work is concerned with GI of the partially coupled gas-dust mixture 
taking into account gas and dust aerodynamical interaction. The linear stability analysis of an unbounded uniform 
self-gravitating medium is carried out in the two-fluid approximation with dust assumed to be a pressureless fluid. 
Hence, the objective of this study is to generalise the classical plane wave solution obtained by Jeans 
for the dynamics of two partially coupled fluids. 
It is shown that in this case the gas-dust mixture is unstable at all scales. 
Additionally, the dust is allowed to drift through the gas 
under the action of some external force. In the latter case, this study generalises the
\citetalias{squire-2018_acoustic} model. 
% Неверно для $V>0$, см. профили инкрементов!!! Исправить формулировку - для дрейфа, если джинсовский масштаб 
% больше масштаба исчезновения дозвуковой неуст. HS18, то по-прежнему так, иначе - нет. 
The resonant instability of a new type is found at the (sub-)Jeans scale. As far as the drift velocity is 
sufficiently small (or equal to zero), this instability operates due to the
dust back-reaction on gas arising from the dust self-gravity.
If the drift velocity is higher than some critical value, the instability operates due to the known aerodynamical 
dust back-reaction on gas caused by the bulk drift of the dust subject to external force. In the latter case, 
the instability is more prominent than that of \citetalias{squire-2018_acoustic} for the subsonic drift. 
The growth rate of the new instabilities depends on either the square
root or the cube root of the dust fraction, which is defined by the different critical value of the drift velocity. 
It is stated that the new resonant instabilities can affect the gravitational collapse of various dust-laden objects,
where grains are significantly decoupled from gas.

%Серия ссылок на аналогичную задачу в дисках.
A related problem has been studied in protoplanetary discs in the context of planetesimal formation. 
A dense sub-disc of macroscopic solids having small but non-zero velocity dispersion is embedded in a gas disc, which 
can usually be assumed gravitationally stable. In the absence of aerodynamic drag, which damps the velocity dispersion 
of solids, the sub-disc would be gravitationally stable as well. However, taking into account aerodynamic drag 
makes the sub-disc of solids unstable. The corresponding instability is referred to as the secular GI,
which operates when both the relative motion of gas and dust and the self-gravity of dust are taken into account, 
see \citet{youdin-2005_GI} and \citet{youdin-2011} who addressed the problem without dust back-reaction on gas, 
or \citet{takahashi-2014} and \citet{latter-2017} who took into account the aerodynamic interaction of gas and dust.

\section{Dynamics of self-gravitating dust-laden medium}

%Transonic regime: $V_{||}\to 1$.
%(Essentially) sunsonic regime: $V_{||} \ll 1$.

%Abbreviate heavy sound wave and sound wave distinctly (HSW and SW, respectively).

% Сказать где-то о том, что величина \delta введена удобно, т.к. показывает прямо эволюцию доли пыли вне зависимости 
% от того, как ведет себя плотность газа. 

%This study considers only the subsonic bulk drift of the dust.

\subsection{Two-fluid equations}

Dynamics of the gas-dust mixture can be considered in the two-fluid approximation.
The fluid associated with gas has velocity ${\bf U}_g$, while ${\bf U}_p$ is the velocity of the fluid associated with dust.
The relative velocity of dust with respect to the gas, ${\bf V}\equiv {\bf U}_p -{\bf U}_g$, drives the aerodynamic drag,
which couples the two fluids to each other.
The corresponding equations of motion and mass conservation are the following
\begin{equation}
\label{eq_U_g}
%\begin{aligned}
\frac{\partial {\bf U}_g}{\partial t} + ({\bf U}_g \cdot \nabla){\bf U}_g = 
- \frac{\nabla p}{\rho_g} - \nabla \Phi + \frac{\rho_p}{\rho_g} \frac{\bf V}{t_s},
%\end{aligned}
\end{equation}
\begin{equation}
\label{eq_rho_g}
\frac{\partial \rho_g}{\partial t} + \nabla\cdot (\rho_g{\bf U}_g) = 0
\end{equation}
for gas with mass density $\rho_g$ and
\begin{equation}
\label{eq_U_p}
%\begin{aligned}
\frac{\partial {\bf U}_p}{\partial t} + ({\bf U}_p \cdot \nabla){\bf U}_p = {\bf a} - \nabla \Phi - \frac{{\bf V}}{t_s},
%\end{aligned}
\end{equation}
\begin{equation}
\label{eq_rho_p}
\frac{\partial \rho_p}{\partial t} + \nabla\cdot (\rho_p{\bf U}_p) = 0.
\end{equation}
for dust with mass density $\rho_p$. Dust is subject to an external force, which causes an acceleration ${\bf a}$, 
see equation (\ref{eq_U_p}).
Aerodynamic drag is represented by the last term on the right-hand side (RHS) of both equations (\ref{eq_U_g}) and (\ref{eq_U_p}). 
It is parametrised by the grain's stopping time, $t_s$, which is assumed to be constant in this study. 
Note that the latter assumption is oversimplifying, since the variations of gas density invoke the corresponding 
variations of aerodynamic drag, see \citetalias{squire-2018_acoustic}\footnote{It can be checked that the addition 
of the corresponding terms in equations given below does not invalidate the 
basic estimates and conclusions made for the resonant instabilities.}.
%{\bf also, references for 
%explicit Epstein drag etc - borrow from SH18}.
The gas pressure is denoted by $p$, while dust is considered to be pressureless. 
New to this work is that the mixture flows in its own gravitational potential, $\Phi$, 
which is determined by its total density %, $\rho \equiv \rho_g + \rho_p$, 
according to Poisson equation
\begin{equation}
\label{Poisson_eq}
\nabla^2 \Phi = 4\pi G (\rho_g + \rho_p).
\end{equation}
In what follows, it is assumed for simplicity that gas is barotropic, $p = p(\rho_g)$, and, accordingly, 
$\nabla p = c_s^2 \nabla \rho_g$, where $c_s$ is the sound speed. 
Thus, the specific pressure gradient in equation (\ref{eq_U_g}) can be replaced by the gradient of a new quantity, 
$\nabla h \equiv \nabla p/\rho_g$, where $h$ is equivalent to enthalpy in the particular case of homentropic flow.
Equations (\ref{eq_U_g}-\ref{Poisson_eq}) specify the dynamics of a self-gravitating partially coupled gas-dust mixture.

\subsection{Stationary self-gravitating configuration}

In order to construct the homogeneous stationary solution, the Jeans swindle is expanded here onto the two-fluid model.
Thus, it is assumed that stationary gravitational potential is zero, while the validity of the Poisson equation 
is guaranteed by some external source of gravity, see e.g. \citet{Binney-1987}.
In this case, the stationary solution obeying equations (\ref{eq_U_g}-\ref{Poisson_eq}) is as follows
\begin{equation}
\label{stat_1}
\Phi = 0,
\end{equation}
\begin{equation}
\label{stat_2}
{\bf U}_g =0, 
\end{equation}
\begin{equation}
\label{stat_3}
{\bf U}_p = {\bf V} = t_s {\bf a},
\end{equation}
\begin{equation}
\label{stat_4}
\frac{\nabla p}{\rho_g} = f\, {\bf a},
\end{equation}
\begin{equation}
\label{stat_5}
\rho_g = const, \quad \rho_p = const,
\end{equation}
where the constant dust fraction is introduced as
\begin{equation}
\label{f}
f\equiv \frac{\rho_p}{\rho_g}.
\end{equation}

Equations (\ref{stat_1}-\ref{stat_5}) represent the homogeneous self-gravitating medium, which consists of gas in
hydrostatic equilibrium and
dust drifting through the gas if it is externally forced.  

%{\bf Now, plug this solution into the general condition: $\lambda^{-1} = f V k t_s$, the shortest characteristic time 
%of the problem is $t_{ev} = c_s k$ --- so that it must be $V k <<1$ and $k<<1$ for the dimensionless quantities. 
%So, it is possible that $V\gtrsim 1$!.. }
%
%Thus, the assumptions (\ref{tau_ev}) and (\ref{l_ev}) are commonly valid is the clouds of cold atomic hydrogen.
%Indeed, ...

\subsection{Equations for linear gas-dust perturbations}

Let the Eulerian perturbations of enthalpy and dust density be denoted as $h^\prime$ and $\rho^\prime_{p}$, respectively,
while the Eulerian perturbation of gas density be 
\begin{equation}
\label{rho_g}
\rho^\prime_{g} = \rho_g \frac{h^\prime}{c_s^2}.
\end{equation}
The Eulerian perturbation of the dust fraction reads
\begin{equation}
\label{f_1} \nonumber
f^\prime \equiv \frac{\rho^\prime_{p}}{\rho_g} - f \frac{\rho^\prime_{g}}{\rho_g}.
\end{equation}
As the both fluids are compressible, the relative perturbation of the dust fraction becomes a meaningful quantity
expressed as
\begin{equation}
\label{delta}
\delta \equiv \frac{f^\prime}{f} = \frac{\rho^\prime_{p}}{\rho_p} - \frac{h^\prime}{c_s^2}.
\end{equation}
Note that as $c_s\to\infty$, $\delta$ tends to the relative perturbation of dust density used, e.g., by 
\citet{zhuravlev-2019}.
Further, the Eulerian perturbations of gravitational potential, gas and dust velocities are, 
respectively, $\Phi^\prime$, ${\bf u}_g$ and ${\bf u}_p$. 
The Eulerian perturbation of the relative velocity is 
\begin{equation}
\label{v}
{\bf v} = {\bf u}_p - {\bf u}_g.
\end{equation}

Equations (\ref{eq_U_g}-\ref{Poisson_eq}) linearised on the background (\ref{stat_1}-\ref{stat_5}) are the following 

\begin{equation}
\label{eq_u_g}
\frac{\partial {\bf u}_g}{\partial t} = -\nabla h^\prime - \nabla \Phi^\prime + f \left ( \frac{{\bf v}}{t_s} + 
\frac{{\bf V}}{t_s} \delta \right ),
\end{equation}

\begin{equation}
\label{eq_cont_g}
\frac{\partial \rho^\prime_g}{\partial t} + \rho_g \nabla\cdot {\bf u}_g = 0,
\end{equation}

\begin{equation}
\label{eq_u_p}
\frac{\partial {\bf u}_p}{\partial t} + ({\bf V}\cdot \nabla) {\bf u}_p = - \frac{{\bf v}}{t_s} - \nabla \Phi^\prime,
\end{equation}

\begin{equation}
\label{eq_cont_p}
\frac{\partial \rho^\prime_p}{\partial t} + ({\bf V}\cdot \nabla) \rho^\prime_p + \rho_p \nabla\cdot {\bf u}_p = 0,
\end{equation}

\begin{equation}
\label{eq_Phi_1}
\nabla^2 \Phi^\prime = 4\pi G \left ( \rho^\prime_g + \rho^\prime_p \right ).
\end{equation}
They are supplemented by equations (\ref{rho_g}), (\ref{delta}) and (\ref{v}).

Taking the divergence of equations (\ref{eq_u_g}) and (\ref{eq_u_p}) and combining equation (\ref{eq_cont_g}) 
with equation (\ref{eq_cont_p}) one arrives at the more compact set of equations

\begin{align}
\label{eq_h_1}
\frac{1}{c_s^2} \frac{\partial^2 h^\prime}{\partial t^2} = \nabla^2 h^\prime + \omega_{ff}^2\left [ (1+f) \frac{h^\prime}{c_s^2} + 
f \delta \right ] - \nonumber \\ 
\frac{f}{t_s} \left [ \nabla \cdot {\bf v} +  ({\bf V}\cdot \nabla) \delta \right ],  
\end{align}
\begin{equation}
\label{eq_delta}
\frac{\partial \delta}{\partial t} + ({\bf V}\cdot\nabla) \delta = - \frac{1}{c_s^2} ({\bf V}\cdot\nabla ) h^\prime 
- \nabla \cdot {\bf v}, 
\end{equation}
\begin{align}
\frac{\partial (\nabla \cdot {\bf v})}{\partial t} - \frac{1}{c_s^2} \frac{\partial^2 h^\prime}{\partial t^2} + 
({\bf V\cdot \nabla}) \left [ ( \nabla\cdot {\bf v} ) - \frac{1}{c_s^2} \frac{\partial h^\prime}{\partial t} \right ] = \nonumber \\
- \frac{1}{t_s} \nabla \cdot {\bf v}
- \omega_{ff}^2 \left [ (1+f) \frac{h^\prime}{c_s^2} + f\delta \right ],
\label{eq_v}
\end{align}
where
\begin{equation}
\label{def_om_ff}
\omega_{ff} \equiv (4\pi G\rho_g)^{1/2}
\end{equation}
is the inverse characteristic free fall time of the gas component, $t_{ff}$.

The set of equations (\ref{eq_h_1}-\ref{eq_v}) is closed with respect to the scalars $h^\prime$, $\delta$ 
and $\nabla\cdot {\bf v}$. It describes the divergence of the velocity fields of dust and gas leaving its vortex components undetermined. 
Gravity does not affect the vortex components of ${\bf u}_{g,p}$ within the model considered in this work. 
The vortical dynamics of the gas-dust mixture is non-trivial due to the aerodynamic drag and, though omitted here, 
is worthy of a separate study. In what follows, ${\bf u}_{g,p}$ are assumed to be the potential fields.
%However, the gravity only affects the velocity divergence, so the vorticity of both ${\bf u}_{g,p}$ 
%is assumed to be absent in the solutions considered in this work. 

%%%%%%%%%%%%%%%%%%%%%%%%%%%%%%%%%%%%%%%%%%%%%%%%%%%%%%%%%

When $f\to 0$, equation (\ref{eq_h_1}) describes the propagation of sound in a self-gravitating gas environment.
The second terms $\propto f$ in the first and the second square brackets on the RHS 
of equation (\ref{eq_h_1}) introduce,
respectively, the gravitational and the aerodynamic feedback from the dust. 
It may seem that these additional terms can hardly change the dynamics of ISM, where typically $f\sim 0.01$, 
however, the case is more interesting as heavy sound waves may come into resonance with the trivial dust mode, see 
Sections \ref{sec_tr_res} and \ref{sec_d_res}.

Equation (\ref{eq_delta}) describes the dynamics of dust in terms of the evolution of the dust to gas density ratio. 
The conservation of the perturbed dust fraction along the bulk stream of the dust described 
by the left-hand side (LHS) of this equation 
is violated through the terms on its RHS. The first term on the RHS of equation (\ref{eq_delta}) 
takes into account the change of the dust fraction due to the bulk advection of dust with a certain density 
into the regions with a different density of gas. 
The second term on the RHS of equation (\ref{eq_delta}) introduces the change of the dust fraction due to the 
divergence of the perturbed dust motion with respect to the gas. 
It is this latter term that provides the clumping of dust in the domains of high pressure 
widely studied in protoplanetary discs.
It is determined by equation (\ref{eq_v}), which is essentially a different form of equation
(\ref{eq_u_p}). %being just an equation of motion for grains. 

Accordingly, equation (\ref{eq_v}) can be thought of as an equation of the perturbed motion of grains considered in 
the frame comoving with the perturbed gas. In this case, the terms $\sim h^\prime$ on the LHS of this equation 
represent the inertial force acting on grains in this frame. The acceleration of grains with respect to gas is
expressed by the rest of the terms on the LHS of equation (\ref{eq_v}), while the terms on its RHS describe 
the aerodynamic drag and the gravitational force. In the limit of the long evolution time of the mixture, 
\begin{equation}
\label{TVA_1}
t_{ev} \gg t_s,
\end{equation}
as well as
the long length-scale of perturbations, 
\begin{equation}
\label{TVA_2}
l_{ev} \gg c_s t_s,
\end{equation}
the second term on the LHS of equation (\ref{eq_v}) becomes
the leading one. It represents the effective gravity acting on grains due to the acceleration of the gas. 
When applied to the non-self-gravitating gas-dust mixture, the limit (\ref{TVA_1}-\ref{TVA_2}) 
is known as the terminal velocity approximation \citep{youdin-goodman-2005, latter-2011}: 
the aerodynamic drag is balanced by the pressure gradient. As the restrictions (\ref{TVA_1}-\ref{TVA_2}) are valid, 
the relative velocity of the dust in the perturbed motion, ${\bf v}$, is not difficult to exclude from equations 
(\ref{eq_h_1}-\ref{eq_v}). However, the analytical results describing the resonant instabilities can 
be obtained with no restrictions on $t_{ev}$ and $l_{ev}$, see Sections \ref{sec_tr_res} and \ref{sec_d_res}.

\subsection{Units}

For the problem considered in this study, the natural units to measure time and velocity are, respectively, $t_s$ and $c_s$.
Accordingly, the length has to be measured in units of $c_s t_s$. 
In the dimensionless version of equations (\ref{eq_h_1}-\ref{eq_v}) the matter's own gravity is introduced by 
the dimensionless parameter
\begin{equation}
\label{tau}
\tau \equiv \omega_{ff} t_s,
\end{equation}
which provides the ratio of stopping time to free-fall time. 
This parameter takes a wide range of values less than unity in molecular clouds, see Section \ref{sec_physics}.

\begin{figure*}%{h}
\begin{center}
\includegraphics[width=18cm,angle=0]{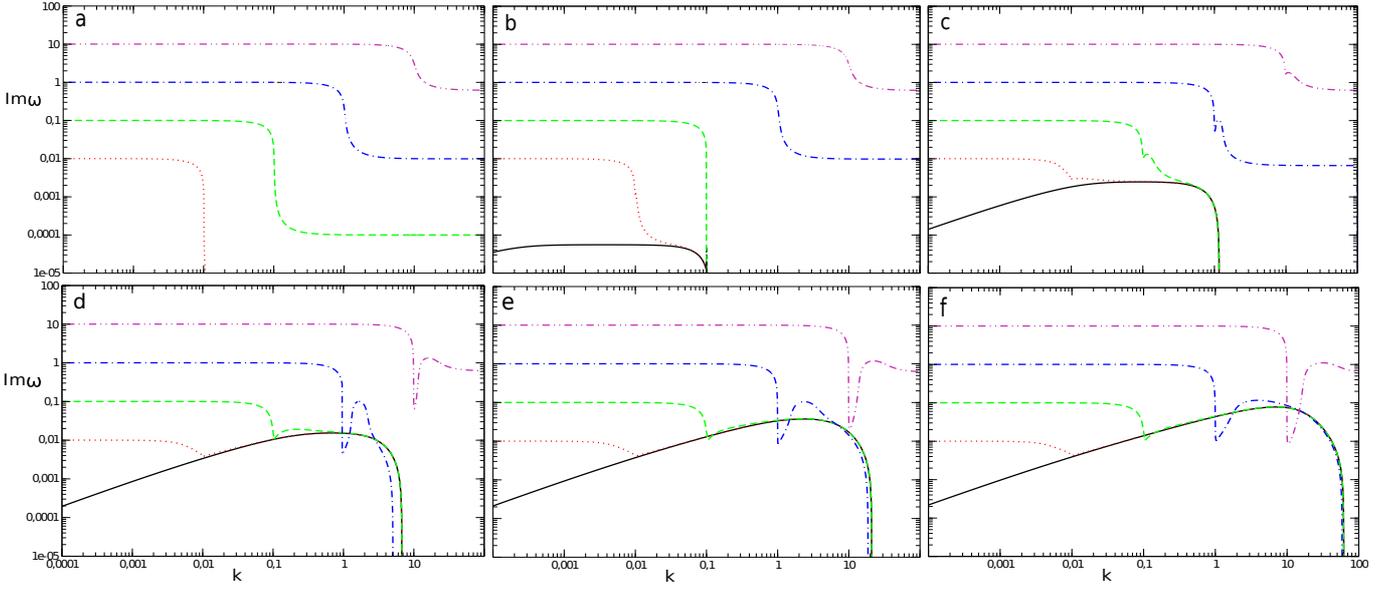}%Im_om_1}
\end{center}
\caption{The largest $\Im[\omega]>0$ among the modes being the solution of equation (\ref{disp}) vs. 
the absolute value of wavenumber for $f=0.01$. $\hat V_{||}=0.0,0.1,0.5,0.8,0.9,0.95$ in 
panels (a), (b), (c), (d), (e), (f), respectively. 
Solid (black), dotted (red), dashed (green), dot-dashed (blue) and dot-dot-dashed (magenta) lines show 
$\tau=0,0.01,0.1,1.0,10$, respectively.
} \label{fig_1}
\end{figure*}

\subsection{Dispersion equation}

The particular solution in the form of a wave with the dimensionless complex frequency $\omega$ 
measured in units of $t_s^{-1}$ and wavevector ${\bf k}$ measured in units of $(c_s t_s)^{-1}$ reads
\begin{equation}
\label{fourier}
h^\prime,\, \delta,\, \nabla \cdot{\bf v} = \{ \bar h^\prime,\, \bar \delta,\, \nabla\cdot {\bf \bar v} \} \,
\exp( -{\rm i} \omega t_* + {\rm i} {\bf k}\cdot {\bf r}_* ),
\end{equation}
where the dimensionless time and length are, respectively,
$$
t_* \equiv \frac{t}{t_s},
$$
$$
{\bf r}_* \equiv \frac{{\bf r}}{c_s t_s}.
$$
Equations (\ref{eq_h_1}-\ref{eq_v}) then yield the dispersion equation, which can be expressed as
\begin{equation}
\label{disp}
D_g(\omega,{\bf k}) \cdot D_p(\omega, {\bf k}) = \epsilon(\omega, {\bf k}),
\end{equation}
with
\begin{equation}
\label{D_g}
D_g(\omega,{\bf k}) \equiv \omega^2 - k^2 + \tau^2 (1+f) + f \,\frac{\omega(\omega-{\bf k}\cdot\hat{\bf V}) + \tau^2 (1+f)}
{1-{\rm i}\omega + {\rm i} {\bf k}\cdot\hat {\bf V}},
\end{equation}
\begin{equation}
\label{D_p}
D_p(\omega,{\bf k}) \equiv \omega - {\bf k}\cdot \hat {\bf V} - \frac{{\rm i f \tau^2}}
{1-{\rm i} \omega + {\rm i} {\bf k}\cdot \hat {\bf V} },
\end{equation}
and
\begin{align}
\epsilon(\omega, {\bf k}) \equiv  f \left [ {\bf k}\cdot\hat {\bf V} + 
\frac{{\rm i}\omega(\omega-{\bf k}\cdot\hat{\bf V}) + {\rm i} \tau^2 (1+f)}
{1-{\rm i}\omega + {\rm i} {\bf k}\cdot\hat {\bf V}} \right ] \nonumber \\ 
\left [ {\rm i} {\bf k}\cdot\hat{\bf V} - \tau^2 \left ( 1 + \frac{f}{1-{\rm i}\omega + {\rm i}{\bf k}\cdot\hat{\bf V}} \right ) 
\right ] ,
\label{eps}
\end{align}
where $\hat {\bf V}$ is the drift velocity of dust measured in units of $c_s$. It is assumed that $\hat V<1$ hereafter.

Equation (\ref{disp}) accurately determines the linear modes of gas-dust perturbations on the homogeneous background 
specified by equations (\ref{stat_1}-\ref{stat_5}). These modes constitute oscillations in the gas-dust medium at the 
frequency given by the real part of $\omega$. The modes can be damping or growing with the amplitude changing
exponentially with time at a rate given by the imaginary part of $\omega$. 
The parameters $f$ and $\tau$ can take any finite values.
However, in the ISM commonly $f\ll 1$. In this case, the solution of equation (\ref{disp}) must be 
close to the basic one corresponding to $f=0$. Strictly at $f=0$, equation (\ref{disp}) splits into two independent equations
\begin{equation}
\label{D_g}
\omega^2 = k^2 - \tau^2 \equiv \omega_s^2 
\end{equation}
and
\begin{equation}
\label{D_p}
\omega = {\bf k}\cdot \hat {\bf V}  \equiv \omega_p.
\end{equation}
Equations (\ref{D_g}) and (\ref{D_p}) describe, respectively, 
perturbations existing in gas and 
a trivial dust mode associated with perturbations of the dust density.

Equation (\ref{D_g}) describes the Jeans instability. 
As far as $k < \tau$, the corresponding modes are the two static waves 
growing and damping at a rate $|\omega_s|$, where $k$ is the absolute value of the mode wavenumber.
In contrast, as $k > \tau$, the modes become two waves propagating in the opposite directions, which are 
the heavy sound waves (HSW hereafter).
The marginal value $k=\tau \equiv k_J$ specifies the Jeans length.

Equation (\ref{D_p}) introduces a wave of perturbations of the dust density advected by the bulk drift of the dust.
Indeed, this equation can be obtained from equations (\ref{eq_h_1}-\ref{eq_v}) in the limit $f\to 0$ 
provided that additionally $h^\prime \to 0$, which implies that $\delta$ tends to the relative perturbation of the dust
density. This wave is essentially the one introduced by \citet{zhuravlev-2019} for the description of the instability
of a gas-dust mixture in protoplanetary discs and referred to as the streaming dust wave (SDW hereafter). 
This term will be used in this study. 
The SDW phase velocity equals the bulk drift velocity of the dust projected onto ${\bf k}$. 
In the particular case of the suspended dust, i.e. $\hat {\bf V}=0$, SDW degenerates into static perturbations 
of the dust density. In this study, it is considered along with the general case $\hat V > 0$.

As the dust fraction acquires a small but non-zero value, the modes of gas-dust perturbations deviate from the solutions
of equations (\ref{D_g}) and (\ref{D_p}) discussed just above.
Foremost, this occurs due to the RHS of equation (\ref{disp}), which will be referred to as the coupling term hereafter.
Additionally, there are corrections to the LHS of equations (\ref{D_g}) and (\ref{D_p}) proportional to $f$.
Such modes will be referred to as the modes akin to HSW or SDW. 
The rest of the paper deals with an accurate numerical solution of equation (\ref{disp}) followed by the analytical
consideration of the particular situations caused by the resonance between HSW and SDW, which takes place at the 
mode crossing.

\begin{figure*}%{h}
\begin{center}
\includegraphics[width=14cm,angle=0]{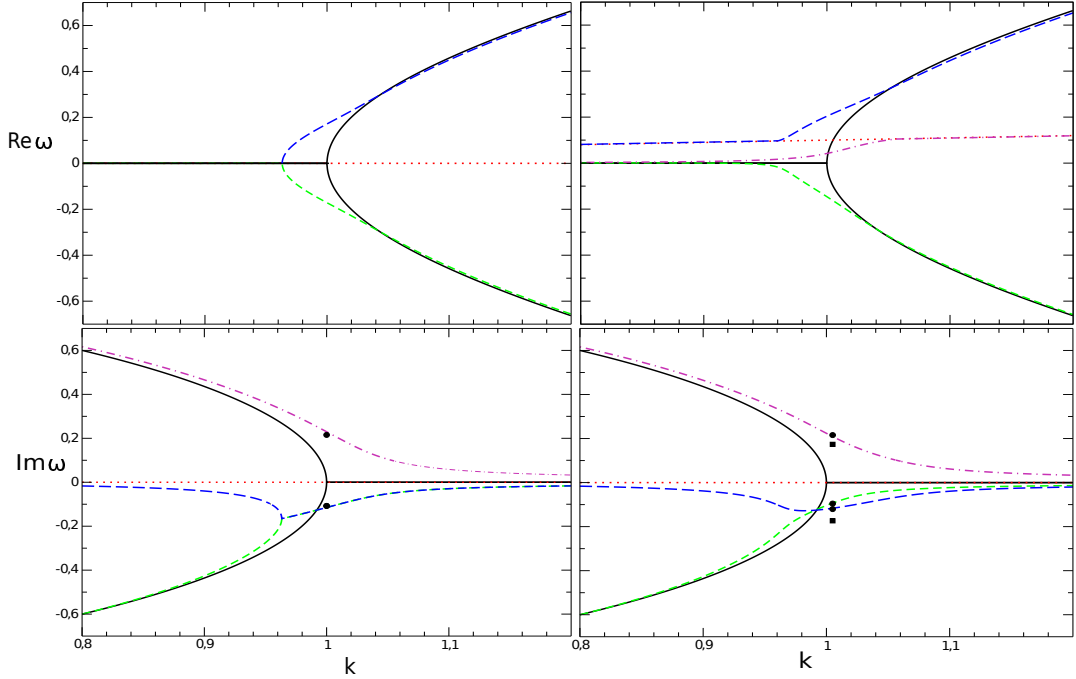}%om_V00}
\end{center}
\caption{The curves in top and bottom panels show, respectively, $\Re [\omega]$ and $\Im [\omega]$, 
where $\omega$ is the solution of equation ({\ref{disp}}). 
Solid (black) and dotted (red) lines show, respectively, two HSW and SDW obtained for $f=0$. 
Short-dashed (green) and long-dashed (blue)
lines show two damping modes, while the dot-dashed (magenta) lines show the growing mode for $f=0.01$. 
The dot-dashed (magenta) line in the top-left panel is not shown as it virtually coincides with the dotted (red) line. 
$\hat V_{||}=0.0$ and $\hat V_{||}=0.1$ in the left and in the right panels, respectively. For all panels $\tau=1.0$.
The filled circles in the bottom-left and bottom-right panels represent the analytical solutions given by equations (\ref{tr_res_simple}) and (\ref{tr_res}), respectively.
The filled squares represent the analytical solution given by equation (\ref{d_res}).
}
\label{fig_2}%fig_4}
\end{figure*}

\section{Profiles of the growth rate}
\label{sec_profiles}

The goal of this section is to reproduce the overall picture of GI of a uniform dust-laden medium 
under the interplay of two effects: the contribution of the dust to the gravitational potential and the 
bulk drift of the dust with respect to the gas.
For that, equation (\ref{disp}) is numerically solved as an algebraic quintic equation with respect to $\omega$.
According to equation (\ref{fourier}), the growth rate of the gas-dust mode is represented by $\Im [\omega]>0$.
The largest growth rate vs. the absolute value of wavenumber is shown in Figure 
\ref{fig_1} for the usual value of the dust fraction, $f=0.01$ and various $\tau$. 

The very first panel, see Figure \ref{fig_1}, represents the special
case of the dust suspended in the gas, i.e. no bulk drift of the dust, ${\bf V}=0$. Note that physically this means the 
absence of an external force acting on grains, rather than the rigid coupling of the grains with the gas, since 
$t_s$ is implied to be a finite unit of time throughout this study. Thus, the relative motion of the grains 
with respect to the gas occurs in the perturbed flow. In this case, the self-gravitating medium becomes unstable 
at all length-scales in contrast to the well-known Jeans solution. 
Perturbations with length-scale equal and smaller than the Jeans length,
$k > k_J$, are no longer stable. This occurs irrespective of the value of $\tau$, 
though the growth rate in the limit of high $k$ decreases visually like $\tau^2$ as one proceeds 
to smaller grains / weaker gravity. For a particular $\tau$, the growth rate
approaches that of the Jeans instability at $k\to 0$, but retains a considerable non-zero value as
$k\to k_J$, where the Jeans instability ceases. In the limit $k\to \infty$ it reaches a smaller horizontal asymptotics.

The rest of the panels in Figure \ref{fig_1} represent how the profiles of the growth rate change as the bulk
drift of the dust progressively increases. The growth rate in each panel in Figure \ref{fig_1} 
is defined by the projection of $\hat {\bf V}$ onto ${\bf k}$, which is denoted as $\hat V_{||}$, 
so that ${\bf k}\cdot \hat {\bf V} \equiv \hat V_{||} k$. Hereafter, $\hat V_{||}$ will be referred to as simply 
the drift velocity. It will be assumed to be a positive value.

As the drift velocity is highly subsonic, the instability of 
\citetalias{squire-2018_acoustic} corresponding to their case of $t_s=const$ emerges in the range of small wavenumbers,
$k\ll 1$, see the solid line in panel (b) in Figure \ref{fig_1} corresponding to $\tau=0$. 
As was discussed by \citetalias{squire-2018_acoustic}, this instability appears when the dust drifts through the
pressureless environment (e.g. cold gas) subject to the aerodynamic back-reaction of dust.
The physics of this instability is discussed here in the Appendix \ref{sec_HS18}.
As long as $\hat V_{||} \ll 1$, the \citetalias{squire-2018_acoustic} instability is weak, so its long-wavelength
side overlaps the Jeans instability already at the very small $\tau$. However, it dominates at $k > k_J$
for $\tau \lesssim \hat V_{||}$. Accordingly, the growth of perturbations ceases at high $k$ for 
$\tau \lesssim \hat V_{||}$. It should be also noted for this case (see the dotted curve in panel (b)
in Figure \ref{fig_1}) that the growth rate at $k \simeq k_J$ has increased as compared to the case $\hat V_{||}=0$.
On the other hand, the growth of perturbations for $\tau \gtrsim \hat V_{||}$ including the case of $\tau >1$ 
remains unchanged at all scales. 

%\begin{figure}%{h}
%\begin{center}
%\includegraphics[width=8cm,angle=0]{figures/Im_om_2}
%\end{center}
%\caption{The same as in Figure \ref{fig_1} but for $\hat V_{||} = 0.5$ and $\hat V_{||}=0.8$ on top and bottom panels, %respectively.
%} \label{fig_2}
%\end{figure}

For a drift velocity approaching the speed of sound, modification of the growth rate profile occurs at all $\tau$.
Additional bumps on profiles of the growth rate can be seen in panels (c) and (d) in Figure \ref{fig_1}. 
These bumps are located at 
sub-Jeans scales, $k \gtrsim k_J$. While the dust drift becomes transonic, $\hat V_{||} \to 1$, 
the bumps shift slowly to higher wavenumbers as compared to $k_J$, compare panels (c), (d), (e) and (f) 
in Figure \ref{fig_1} with each other. At the same time, dips in the growth rate 
on scales between that of the corresponding bumps and the Jeans scale get slightly wider. The most 
prominent bumps are found for $\tau \sim 1$, see the particular case in panel (d) in Figure \ref{fig_1}.
The bumps that emerge at small $\tau \ll 1$ approximately follow the enhanced profile of 
\citetalias{squire-2018_acoustic} instability for the dust drift getting transonic (see the particular case
of $\tau=0.1$ in panels (d), (e), (f) in Figure \ref{fig_1}). 
This is so even though they exceed the \citetalias{squire-2018_acoustic} instability growth rates as $\hat V_{||}$ is 
far from unity (see $\tau=0.1$ in panel (c) in Figure \ref{fig_1}).
At the same time, the bumps found at $\tau \sim 1$ exhibit a substantially higher growth rate than that of 
the \citetalias{squire-2018_acoustic} instability up to $\hat V_{||}$ very close to unity (see the particular case
of $\tau=1.0$ in panel (f) in Figure \ref{fig_1}). On the other hand, the bumps corresponding 
to $\tau \gg 1$ hardly exceed the growth rate obtained in the absence of the dust drift at the same scales regardless 
of the value of $\hat V_{||}$. Instead, the dips make the main difference
to the growth rate profiles for $\tau \gg 1$ as compared to the case of $\hat V_{||} = 0$, 
see panels (e) and (f) in Figure \ref{fig_1}.

%{\bf Note that the growth rate with the account of self-gravity goes below the curve of the instability of 
%\citetalias{squire-2018_acoustic}.}

%\begin{figure}%{h}
%\begin{center}
%\includegraphics[width=8cm,angle=0]{figures/Im_om_3}
%\end{center}
%\caption{The same as in Figure \ref{fig_1} but for $\hat V_{||} = 0.9$ and $\hat V_{||}=0.95$ on top and bottom panels, %respectively.
%} \label{fig_3}
%\end{figure}

\section{Description of modes}

The particular case of $\tau=1.0$ is adopted in this section to show the behaviour 
of the three essential modes of gas-dust perturbations in detail. These modes are
the two oppositely propagating HSW and SDW in the limit of negligible dust fraction. In the same limit, the remaining 
two roots of equation (\ref{disp}) are identical to each other representing trivial modes associated with 
the arbitrary relative motion of dust with respect to gas. 
This motion is damped due to aerodynamic drag, thus $\Im[\omega]\to -1$,
and advected by the bulk drift of the dust, thus $\Re[\omega]\to k \hat V_{||}$. 
Note that $\Re[\omega]$ is referred to as the frequency of the mode hereafter, 
which is not to be confused with the complex frequency in the particular solution (\ref{fourier}).
For finite $f>0$, the trivial modes are modified by gas dynamics, which, in turn, is induced by the dust back-reaction 
on the gas. It is checked that the non-zero $f\ll 1$ makes the trivial modes slightly different from each other, 
however, they always remain damped and close to the basic solution for $f\to 0$. 

At first, each Figure from \ref{fig_2} up to \ref{fig_5} shows the frequencies and the growth (damping) rates of HSW having
positive and negative frequencies at $k>k_J$ and SDW as the solutions of equation (\ref{disp}) for $f=0$. 
Also, these Figures show the modes of gas-dust perturbations akin to HSW and SDW 
as the solutions of equation (\ref{disp}) for $f>0$.

\subsection{GI on account of suspended dust}
\label{sec_suspend_dust}

In the absence of the dust drift in the background solution, SDW exists in the form of static perturbations of 
the dust fraction, thus, it is represented by the zero frequency and growth rate, see the left panels in 
Figure \ref{fig_2}. 
SDW is crossed by either of the two HSW branches at the same point $k=k_J$, see the top-left panel in Figure \ref{fig_2}, 
where the both their frequency and their growth (damping) rate vanish as well. As soon as the dust fraction takes a
finite value, there are the modes of gas-dust perturbations, %akin to HSW are 
which are the slowly propagating damping waves at $k=k_J$. For $k>k_J$, these modes acquire 
the increasing opposite frequencies approaching the frequencies of HSW, 
though their damping rate decreases. On the contrary, at the scales longer than some 
scale corresponding to $k<k_J$, these modes turn into static waves damping at two different rates.
At the same time, the third mode represents a static wave growing at all wavenumbers, i.e. 
its frequency is always equal to zero.
Note that for the particular $\tau=1.0$ and $f=0.01$ the growth rate of GI of dust-laden medium at the Jeans length-scale
is slightly higher than $0.2$, which is a considerable fraction of the free fall rate equal to unity. 
An intuitive description of dust-gas dynamics leading to such a considerable instability is given in the Appendix \ref{sec_Jw}, which additionally should be compared with long- and short-wavelength limits discussed in 
\ref{sec_lw} and \ref{sec_sw}.

There is an ambiguity in the classification of modes for non-zero $f$, 
depending on what quantity is decisive: the frequency or the growth rate. Let the frequency of mode be such a quantity 
in this study. If so, the two damping modes should be classified hereafter as the modes akin to HSW, while the third mode
having the zero frequency should be classified as the one akin to SDW. 
Clearly, the latter mode represents the instability plotted in panel (a) in Figure \ref{fig_1}.
It has been already noted in Section \ref{sec_profiles} that, as $k\to 0$, the growth rate of the mode termed as the one 
akin to SDW approaches that of HSW producing the Jeans instability.

\subsection{GI on account of streaming dust}
\label{sec_stream_dust}

%\begin{figure}%{h}
%\begin{center}
%\includegraphics[width=8cm,angle=0]{figures/om_V01}
%\end{center}
%\caption{The same as in Figure \ref{fig_4} but for $\hat V_{||}=0.1$. Not like in Figure \ref{fig_4}, 
%the dot-dashed line is shown on top panel as it differs from the dotted line.
%The filled circles represent the analytical solution given by equation (\ref{tr_res}).
%The filled squares represent the analytical solution given by equation (\ref{d_res}).
%}
%\label{fig_5}
%\end{figure}

Introduction of the dust drift in the background solution complicates the wave pattern, see the right panels in 
Figure \ref{fig_2}. 
SDW acquires the non-zero positive frequency and crosses the positive frequency HSW at $k>k_J$.
It is seen that the negative frequency mode akin to HSW takes a small negative frequency at all $k<k_J$.
More noteworthy, the positive frequency mode akin to HSW acquires the non-zero frequency, i.e. becomes a
propagating wave, at all $k<k_J$ because it tends to
the frequency of SDW as $k\to 0$. On the other hand, the frequency of the mode akin to SDW retains its zero asymptotics at
$k\to 0$, while it tends to the frequency of SDW itself at high wavenumbers, i.e. $\Re[\omega] \to \hat V_{||} k$ for this
mode. Thus, the growing gas-dust perturbations are represented by a propagating wave now.
It is also notable that the mode crossing of the positive frequency HSW and SDW in the case $f\to 0$ is replaced 
by what is referred to as the avoided crossing of modes akin to HSW and SDW. 
The avoided crossing of modes is the effect known from the theory of waves in plasma and (single-)fluid mechanics, 
see e.g. the book by \citet{stepanyants-fabrikant-1998}. 
% Здесь сделать связку с комментарием к записи общей дисперсионки - с участием coupling term...
At the same time, the degeneracy of modes akin to HSW over the damping rate at $k \gtrsim k_J$ is removed, 
whereas the growth rate of the mode akin to SDW exhibits no visual changes, see the bottom-right panel 
in Figure \ref{fig_2}.
Additionally, it can be checked that the positive frequency mode akin to HSW becomes growing 
in the limit $k \to 0$\footnote{Not seen in the Figure \ref{fig_2}.}.

\begin{figure}%{h}
\begin{center}
\includegraphics[width=8cm,angle=0]{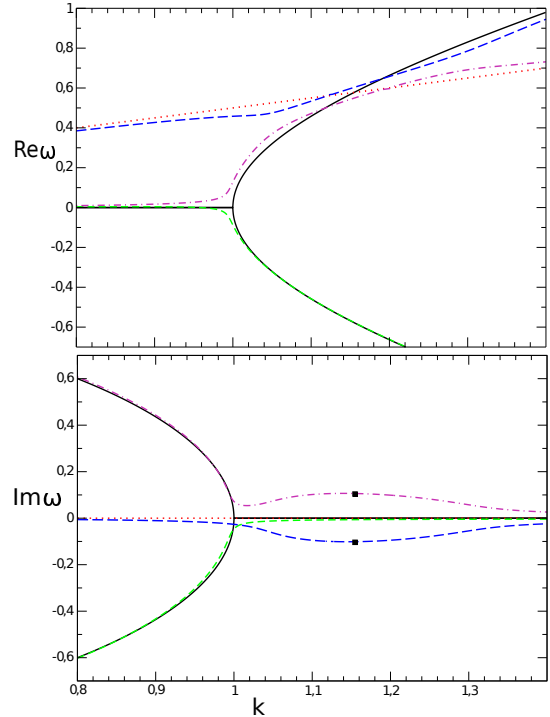}
\end{center}
\caption{The same as in the right panels in Figure \ref{fig_2} for $\hat V_{||}=0.5$.
The filled squares represent the analytical solution given by equation (\ref{d_res}).}
\label{fig_3}%fig_6}
\end{figure}

\begin{figure*}%{h}
\begin{center}
\includegraphics[width=14cm,angle=0]{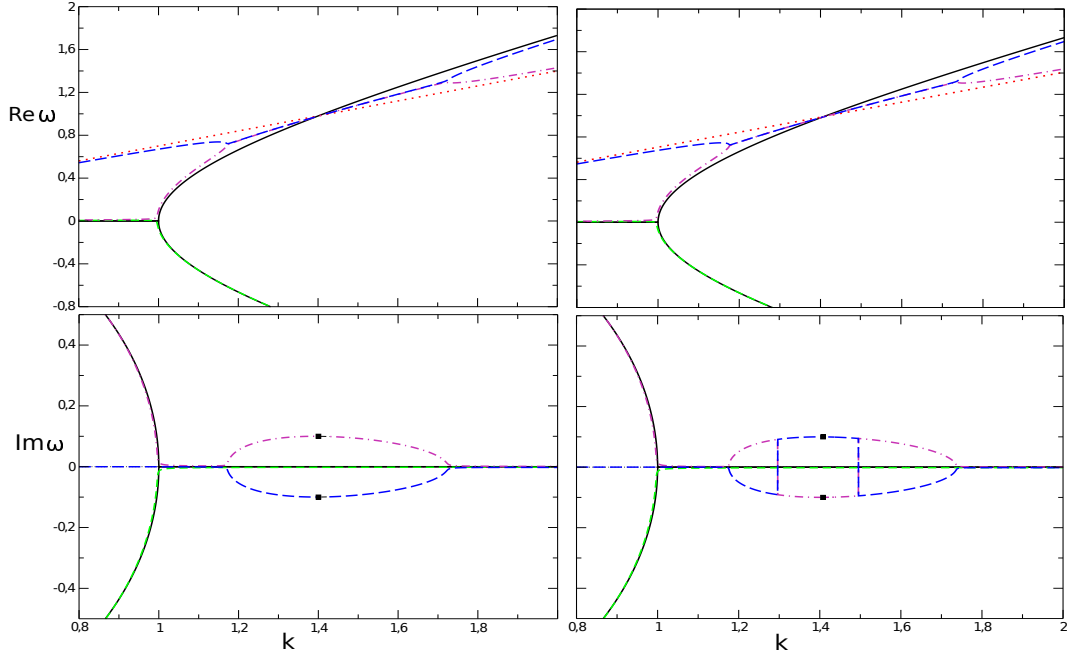}%om_V07}
\end{center}
\caption{The same as in Figure \ref{fig_2} for $\hat V_{||}=0.7$ and $\hat V_{||}=0.704$ 
in the left and in the right panels, respectively.
There is a band where growing and damping modes flip over in the bottom-right panel.
The filled squares represent the analytical solution given by equation (\ref{d_res}).}
\label{fig_4}%fig_7}
\end{figure*}

The higher drift velocity shifts the mode crossing of positive frequency HSW and SDW away from the Jeans length-scale to 
smaller scales as can be seen by comparing the top-right panel in Figure \ref{fig_2} with 
the top panel in Figure \ref{fig_3}. 
This is accompanied by the appearance of a
bump in the growth rate of the mode akin to SDW with the largest growth rate attained approximately at the scale of the 
mode crossing, see the bottom panel in Figure \ref{fig_3}. That is the bump found in Figure \ref{fig_1}.
Clearly, the corresponding growing perturbations are represented by a wave propagating approximately with 
the drift velocity projected onto the wavevector.
The largest damping rate of the positive frequency HSW is shifted from $k \simeq k_J$ to the scale of the 
mode crossing as well. Therefore, the mode crossing considered here provides the enhancement of 
both growth and damping rates of modes akin to those taking part in the mode crossing. 
Specifically for $\tau=1.0$, the value of this enhancement is slightly less than the growth 
rate of self-gravitating dust-laden medium in the absence of the dust drift: compare the left-bottom panel in Figure 
\ref{fig_2} (also the right-bottom panel over there) and the bottom panel in Figure \ref{fig_3}. At the same time, the growth rate of the mode akin to SDW taken 
at the Jeans scale decreases as compared to the case of $\hat V_{||}=0$.
Additionally, it can be checked that the growth of the positive frequency mode akin to HSW gradually expands 
from $k \to 0$ to smaller scales as the drift velocity increases\footnote{Not seen in the Figure \ref{fig_3}.}.
Finally, there are no qualitative changes in frequency and damping rate profiles of negative frequency HSW, compare
Figures \ref{fig_2} and \ref{fig_3}.

%\begin{figure}%{h}
%\begin{center}
%\includegraphics[width=8cm,angle=0]{figures/om_V0704}
%\end{center}
%\caption{The same as in Figure \ref{fig_5} but for $\hat V_{||}=0.704$. 
%There is a band where growing and damping modes flip over.}
%\label{fig_8}
%\end{figure}

Further approach to the transonic dust drift, $\hat V_{||}\to 1$, causes a new effect. 
As soon as the drift velocity increases up to 
$\hat V_{||} \simeq 0.7$ particularly for $\tau=1.0$, the mode crossing of positive frequency HSW and SDW is replaced  
by what is referred to as the mode coupling rather than the avoided crossing: compare the top-left panel in Figure \ref{fig_4} and the top panel in Figure \ref{fig_3}. The top-left panel in Figure \ref{fig_4} shows that 
the frequencies of modes akin to those 
crossing for $f \to 0$ become identical to each other in some band of wavenumbers around the mode crossing scale as 
the dust fraction is finite. 
This is a known feature of mode coupling, see e.g. the review by \citet{stepanyants-fabrikant-1998} and the 
bibliography referenced by \citet{zhuravlev-2019}. 
The gas-dust mixture is most unstable inside the band of the mode coupling, whereas the growth of perturbations 
outside of this 
band is relatively weak: compare the left panels in Figure \ref{fig_4}.
Hence, perturbations at the Jeans length-scale, as well as in some range of the shorter scales, 
exhibit much weaker growth than at the scale of the mode crossing.

\begin{figure}%{h}
\begin{center}
\includegraphics[width=8cm,angle=0]{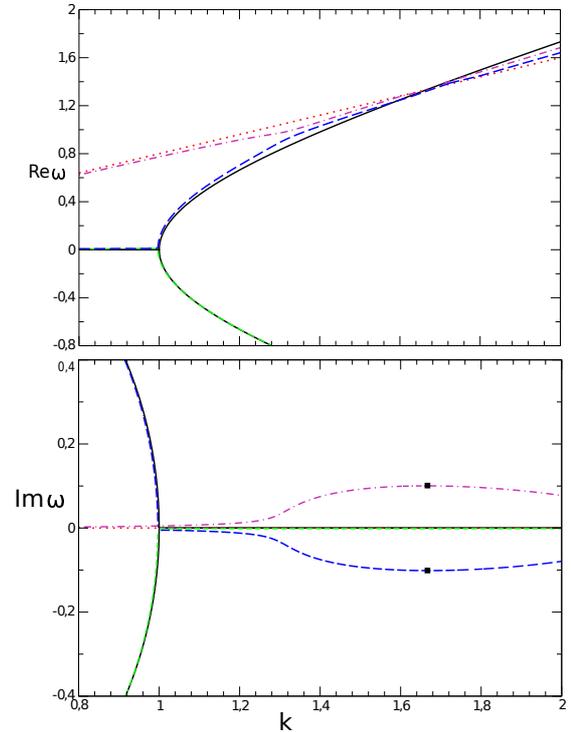}
\end{center}
\caption{The same as in Figure \ref{fig_3} for $\hat V_{||}=0.8$.}
\label{fig_5}%fig_9}
\end{figure}

Additionally, the right panels in Figure \ref{fig_4} presented for the drift velocity close to that in the left panels of 
this Figure show one more effect. 
This is the interchange of modes akin to those participating in the mode crossing by the
growth/damping rates. It can be seen that as the drift velocity takes a tiny increase from $\hat V_{||}=0.7$ to 
$\hat V_{||}=0.704$ particularly for $\tau=1.0$, the additional range of scales within the band of the mode coupling
appears, where the damping mode turns into the growing one and vice versa, see the bottom-right panel in Figure \ref{fig_4}.
This interchange by the growth/damping rate always emerges around the crossing scale and expands to both
small and large scale as the drift velocity increases. The mode interchange stops as it attains approximately 
$k \simeq k_J$ and one finds the new pattern of modes after the mode coupling is transformed to the avoided crossing 
back again, which is demonstrated in Figure \ref{fig_5} for $\hat V_{||}=0.8$. 
The frequency of the growing mode, which is referred to as the mode akin to SDW as before, has now a different asymptotics
far from the mode crossing. As $k\to 0$ and $k\to \infty$, it approaches the frequencies of SDW and  
the positive frequency HSW, respectively, see the top panel in Figure \ref{fig_5}. 
The growth rate of this mode ceases rather than approaching the free fall rate for $k \to 0$: compare the bottom panels in Figures \ref{fig_5} and \ref{fig_3}. The second mode participating in the avoided crossing in Figure \ref{fig_5} approaches
both the frequency and the growth rate of HSW as $k\to 0$, while it becomes damping at
sufficiently high wavenumbers. The third mode akin to the negative frequency HSW has a form qualitatively similar 
to that before and remains damping at all wavenumbers: compare Figures \ref{fig_5} and \ref{fig_3}.

While the drift velocity enters the transonic regime, $\hat V_{||} \to 1$, the band of instability corresponding 
to the avoided crossing\footnote{which is restored after the mode coupling} widens. It eventually recovers the growth rate 
produced by \citetalias{squire-2018_acoustic} instability. It can be noticed that, as the drift velocity tends to unity, 
the growth rate profile of \citetalias{squire-2018_acoustic} instability rises up to the value of the growth rate found 
for self-gravitating dust-laden medium at the scale of the mode crossing long before the transonic regime.

\section{Resonance of modes}

This Section is focused on the analytical approach to the new instabilities found in the vicinity of mode crossings, 
see the previous Section. 
In the vicinity of the mode crossing defined by the condition
\begin{equation}
\label{res_cond}
\omega_s = \omega_p \equiv \omega_c,
\end{equation}
equation (\ref{disp}) can be written in a reduced form corresponding to the leading order in small $f \ll 1$:
\begin{equation}
\label{res_disp}
(\omega - \omega_s) (\omega + \omega_s) (\omega - \omega_p) = \epsilon_c,
\end{equation}
where the reduced form of the coupling term is
\begin{equation}
\label{eps_res}
\epsilon_c \equiv \epsilon({\bf k}, \omega=\omega_c) = - {\rm i} f (\tau^2 - {\rm i} \omega_c)^2.
\end{equation}

Equation (\ref{res_disp}) is obtained by omitting the corrections $\sim f$ on the LHS of equation (\ref{disp}), 
which become of the order higher than $O(f)$ near $\omega_c$, and neglecting the change of the coupling term due to the 
small deviation of $\omega$ from $\omega_c$.
Equation (\ref{res_disp}) can be solved accurately as a cubic equation with respect to $\omega$, however, a further
approximation can be done employing the smallness of the coupling term, $\epsilon_c$.

Let the deviation from the mode crossing frequency caused by the non-zero coupling term be 
$\Delta = \omega - \omega_c$.
As far as 

\begin{equation}
\label{Delta_gg_om}
|\Delta| \gg |2\omega_s|,
\end{equation}
$\Delta$ is determined by equation
\begin{equation}
\label{Delta_triple}
\Delta \approx \epsilon_c^{1/3}.
\end{equation}

In the opposite case, when
\begin{equation}
\label{Delta_ll_om}
|\Delta| \ll |2\omega_s|,
\end{equation}
$\Delta$ is determined by equation
\begin{equation}
\label{Delta_double}
\Delta \approx \pm \left ( \frac{\epsilon_c}{2\omega_c} \right )^{1/2}.
\end{equation}

Equations (\ref{Delta_triple}) and (\ref{Delta_double}) represent the emergence of the resonant instabilities 
in the vicinity of mode crossings of three and two modes, respectively. 
As soon as $\tau >0$ along with $\hat V_{||} > 0$, there is always a crossing of two modes, 
which are SDW and HSW propagating along the same direction. 
They come into resonance at any non-zero $f>0$.
However, the third mode being the counter-propagating HSW joins the resonance
provided that it's frequency is close enough to $\omega_c$, or equivalently, the coupling term is sufficiently strong.
Depending on the value of $\hat V$, both regimes of the modal resonance just introduced above 
may occur is astrophysical conditions, where the dust fraction is commonly a small value. 

The location of the mode crossing determined by equation (\ref{res_cond}) yields 
\begin{equation}
\label{omega_c}
\omega_c = k_c \hat V_{||},
\end{equation} 
where
\begin{equation}
\label{k_c}
k_c = \frac{\tau}{(1-\hat V_{||}^2)^{1/2}}.
\end{equation}
As far as the dust drift is deeply subsonic, $\hat V_{||} \ll 1$, $k_c \simeq \tau$. 
This implies that slowly drifting dust tends to cause the resonance of three modes, since the frequency distance between two HSW $2\omega_s \to 0$ as $k_c \to k_J$.
In contrast, the mode crossing shifts to high wavenumbers, $k\to \infty$, as the drift velocity approaches 
the speed of sound and the modes propagate downstream, $\hat V_{||} \to 1$. 
In the latter case the gas-dust dynamics should encounter the resonance of two modes, 
since the distance between the two HSW increases to high wavenumbers.
The following Sections provide estimates of $\Delta$ for the resonances of the both types.

\subsection{Resonance of three modes}
\label{sec_tr_res}

This resonance is denoted as ${\cal R}_3$ hereafter.

Plugging equation (\ref{eps_res}) into equation (\ref{Delta_triple}) one obtains the following three roots:
\begin{align}
\label{tr_res}
\Delta \approx f^{1/3} (\omega_c^2 + \tau^4)^{1/3} \times \hspace{4cm}\nonumber \\ 
\left \{ \exp \left ( {\rm i}\, \frac{2}{3}\arccos \Psi \right ),\, 
\exp \left ( {\rm i}\, \frac{2}{3}\arccos \Psi \pm {\rm i}\, \frac{4\pi}{3} \right ) \right \}, 
\end{align}
where
\begin{equation}
\label{Psi}
\Psi \equiv \frac{\omega_c-\tau^2}{\sqrt{2}(\omega_c^2 + \tau^4)^{1/2}}
\end{equation}
and $\omega_c$ is defined by equation (\ref{omega_c}).
It can be checked that the first root in equation (\ref{tr_res}) gives the largest growth rate 
irrespective of the values of $\tau$ and $\hat V_{||}$.

It is instructive to consider the following limiting cases for equation (\ref{tr_res}).

\begin{itemize}

\item[i)]
The subsonic regime of strong self-gravity, $\hat V_{||}\ll 1$ and $\tau \gg \hat V_{||}$:
\begin{equation}
\label{tr_res_simple}
\Delta \approx \left \{ {\rm i},\, \frac{\pm \sqrt{3} -{\rm i} }{2} \right \} f^{1/3} \tau^{4/3}.
\end{equation}

\item[ii)]
The subsonic regime of weak self-gravity, $\tau \ll \hat V_{||}\ll 1$:
\begin{equation}
\label{tr_res_simple_weak}
\Delta \approx \left \{ - {\rm i},\, \frac{\pm \sqrt{3} + {\rm i} }{2} \right \} f^{1/3} \tau^{2/3} \hat V_{||}^{2/3}.
\end{equation}

\end{itemize} 

Estimates of the growth rate given by equations (\ref{tr_res_simple}) and (\ref{tr_res_simple_weak}) 
match each other at $\hat V_{||}\sim \tau$ up to a factor of the order of unity. 
Thus, the increase of the drift velocity in the regime of weak self-gravity leads to an additional 
increase in the growth rate as compared with the regime of strong self-gravity. 
The low power of $f$ in both of these estimates explain the substantial growth rate of gas-dust perturbations 
in the medium, which would be marginally stable at $k \gtrsim k_J$ in the absence of dust. 
In the regime of strong self-gravity,
which includes the case of no bulk drift of the dust, the growth rate given by 
equation (\ref{tr_res_simple}) measured in the units of $t_{ff}^{-1}$ behaves like $(f\tau)^{1/3}$.
This implies that as $\tau$ exceeds unity, $\tau \to f^{-1}$, the dust drives clumping of matter at a rate approaching 
the characteristic inverse free fall time.

Note that the transonic regime, $\hat V_{||} \to 1$, in application to ${\cal R}_3$ is not considered in this study. 
As shown in Section \ref{sec_res_type},  ${\cal R}_3$ is restricted by the range $\hat V_{||} < 1$ 
as long as $\tau$ does not exceed $f^{-1}$, which is unrealistic under the conditions in ISM.

Estimate (\ref{tr_res_simple}) is in a good agreement with an accurate solution of the general equation (\ref{disp}), 
see the bottom-left panel in Figure \ref{fig_2}. It also reproduces the avoided crossing between the gas-dust modes 
seen in the top-left panel in Figure \ref{fig_2}.
Further, the more general estimate (\ref{tr_res}) remains good until the value of $\hat V_{||}$
corresponding to the change of the resonance type, see the bottom-right panel in Figure \ref{fig_2} and Section \ref{sec_res_type}. 
As can be seen in the bottom-right panel in Figure \ref{fig_2}, though equation (\ref{tr_res_simple}) is still valid for parameters used there, the more general equation (\ref{tr_res}) additionally reproduces the difference of 
the damping rates of gas-dust modes. 

The regime represented by equation (\ref{tr_res_simple_weak}) operates at rather small $\tau$ as well as $\hat V_{||}$, 
which are not presented in the numerical results considered above. 
However, it is included into analytical description of the 
general picture of modal resonances, see Section \ref{sec_res_type}.

\subsection{Resonance of two modes}
\label{sec_d_res}

This resonance is denoted as ${\cal R}_2$ hereafter.

In this case, equation (\ref{eps_res}) plugged into equation (\ref{Delta_double}) yields
\begin{equation}
\label{d_res}
\Delta \approx \pm \frac{f^{1/2}}{2} \frac{\omega_c - \tau^2 + {\rm i} (\omega_c + \tau^2)}
{\omega_c^{1/2}},
\end{equation}
where $\omega_c$ is again defined by equation (\ref{omega_c}).

As can be seen in Figures \ref{fig_2}-\ref{fig_5}, equation (\ref{d_res}) provides an excellent estimate of 
the growth rate at the mode crossing located at (\ref{k_c}). A poorer match is seen in the bottom-right panel 
in Figure \ref{fig_2}, where the influence of the third mode on the resonance of positive frequency HSW and SDW 
is still considerable, see Section \ref{sec_res_type}.
Equation (\ref{d_res}) is also in accordance with the properties of the frequencies of gas-dust modes. Indeed, as far as
$\tau^2 \gtrsim \omega_c$, what is shown in the top-right panel in Figure \ref{fig_2} and also in the top panel
in Figure \ref{fig_3}, the gas-dust modes modified by a resonance 
undergo the avoided crossing, and the growing mode passes below the damping mode, i.e. the shift of the 
growing/damping mode from the mode crossing is negative/positive, $\Re[\Delta]<0\,/\,\Re[\Delta]>0$, 
in accordance with equation (\ref{d_res}). 
In contrast, as $\tau^2 \lesssim \omega_c$, this situation is reversed, see Figure \ref{fig_5}. The 
transitional case  $\tau^2 \simeq \omega_c$ is demonstrated in Figure \ref{fig_4}, when the modes 
undergo coupling, i.e. they coalesce with each other giving birth to the coupled modes. 
The coupled modes are represented by a complex conjugate pair of $\omega$. Their
frequencies are identical to each other in accordance with equation (\ref{d_res}). 
The special case of the mode coupling is considered in a more detail in Section \ref{sec_coupl}.

Equation (\ref{d_res}) is more tractable in certain limiting cases.

\begin{itemize}

\item[i)]
The subsonic regime of strong self-gravity, $\hat V_{||} \ll 1$ and $\tau \gg \hat V_{||}$:
\begin{equation}
\label{tau_V_1}
\Delta \approx \pm \frac{f^{1/2}}{2} \frac{\tau^{3/2}}{\hat V_{||}^{1/2}} (-1 + {\rm i}).
\end{equation}

\item[ii)]
The subsonic regime of weak self-gravity, $\tau \ll \hat V_{||} \ll 1$:
\begin{equation}
\label{tau_V_2}
\Delta \approx \pm \frac{f^{1/2}}{2} \tau^{1/2} \hat V_{||}^{1/2} (1 + {\rm i}).
\end{equation}

\item[iii)]
The transonic regime of weak self-gravity, $\hat V_{||} \to 1$ and $\tau \ll \delta^{-1/2}$,
where $\delta \equiv 1 - \hat V_{||} \ll 1$:
\begin{equation}
\label{tau_less_1}
\Delta \approx \pm \frac{f^{1/2}\tau^{1/2}}{2} (2\delta)^{-1/4} (1+{\rm i}). 
\end{equation}

\item[iv)]
The transonic regime of strong self-gravity, $\hat V_{||} \to 1$ and $\tau \gg \delta^{-1/2}$:
\begin{equation}
\label{tau_higher_1}
\Delta \approx \pm \frac{f^{1/2}\tau^{3/2}}{2} (2\delta)^{1/4} (-1+{\rm i}). 
\end{equation}

\end{itemize}

Similarly to the case of ${\cal R}_3$, estimates of the growth rate following from equations 
(\ref{tau_V_1}) and (\ref{tau_V_2}) match each other at $\hat V_{||} \sim \tau$ up to the factor of the order of unity.
The change of approximations from (i) to (ii) corresponds to the lowest growth rate, which is of the order of 
\begin{equation}
\label{min_growth_double_res}
\Delta \simeq {\rm i}\,f^{1/2} \tau.
\end{equation}
The growth rate due to the resonance of HSW and SDW generally exceeds (\ref{min_growth_double_res}) 
for both $\hat V_{||}$ higher and lower than small $\sim \tau$. Thus, equation 
(\ref{min_growth_double_res}) can be used as a simple lower estimate of the growth rate for ${\cal R}_2$.
Note that estimate (\ref{min_growth_double_res}) expressed in the units of $t_{ff}^{-1}$ 
yields simply ${\rm i} f^{1/2}$ being independent of $\tau$, i.e. the grain size. 
As far as $\tau$ is small, the regime of weak self-gravity continuously changes from subsonic (ii) to transonic (iii) variant
with the increase of $\hat V_{||}$. 
As can be seen from equations (\ref{tau_V_2}) and (\ref{tau_less_1}), this leads to a further slow increase 
of the growth rate caused mainly by the shift of the mode crossing to larger $k$.

For $\tau \gtrsim 1$, the value of the drift velocity corresponding to the lowest growth rate approaches unity, 
$\hat V\to 1$.
In this case, as $\hat V_{||}$ increases from the small values, the subsonic regime of strong self-gravity 
(i) is replaced by the transonic regime of strong self-gravity (iv), 
which leads to a further slow decrease of the growth rate, see 
equation (\ref{tau_higher_1}). While $\hat V_{||} \to 1$, $\delta^{-1/2}$ becomes comparable to $\tau$, which causes 
a continuous change to transonic regime of weak self-gravity (iii). Again, the growth rate starts to slowly increase.
Note that the lowest growth rate corresponding to the change between the regimes (iv) and (iii) is also estimated by
equation (\ref{min_growth_double_res}).

As $\hat V_{||}$ goes back to zero, while $\tau$ and $f$ remain constant, ${\cal R}_2$ must be replaced by ${\cal R}_3$, 
see the conditions (\ref{Delta_gg_om}) and (\ref{Delta_ll_om}). This can occur in the regimes of either strong or 
weak self-gravity depending on the value of $\tau$, see Section \ref{sec_res_type} and tables \ref{tab_1}, \ref{tab_2}.

\subsection{Mode coupling}
\label{sec_coupl}

Equation (\ref{d_res}) indicates that once 
\begin{equation}
\label{coupling_cond}
\omega_c = \tau^2,
\end{equation}
$\Delta$ becomes imaginary. Thus, the gas-dust modes akin to 
the positive frequency HSW and SDW exhibit identical frequencies\footnote{and also phase velocities} 
equal to those HSW and SDW themselves have at the mode crossing. 
At the same time, damping and growth rates of modes have the same absolute value
\begin{equation}
\label{coupl_Delta}
\Delta \approx \pm {\rm i} f^{1/2} \tau,
\end{equation}
which recovers an order-of-magnitude lower estimate of the growth rate in ${\cal R}_2$, see
equation (\ref{min_growth_double_res}).
In (single-)fluid dynamics, such modes are referred to as the coupled modes after \citet{cairns-1979} 
who applied the concept of mode coupling to the
explanation of the Kelvin-Helmholtz instability. 
The condition of the mode coupling (\ref{coupling_cond}) can be expressed with respect to drift velocity
\begin{equation}
\label{coupl_cond}
\hat V_{||} = \frac{\tau}{(1+\tau^2)^{1/2}},
\end{equation}
which gives $\hat V \approx 0.7$ for the case $\tau=1.0$ shown in Figure \ref{fig_4}. The accurate
solution shown in Figure \ref{fig_4} demonstrates that the coupled modes exist in some interval around 
the mode crossing producing a distinctive ``bridge'' of instability \citep{glatzel-1988}. 
The simplified model of coupling between HSW and SDW, which leads to the reduced dispersion equation (\ref{res_disp}) with
the real coupling term, 
makes it possible to quantify the energy of modes involved in resonance. As expected for problems of this kind,
HSW and SDW coalescing into the coupled modes have positive and negative energies, respectively, see the
Appendix \ref{sec_mode_energy} for details.
According to the common interpretation of the corresponding instability in such a problems, the growth of perturbations 
is caused by the energy flow from SDW having negative energy to HSW having positive energy. The total energy of 
the system of modes remains unchanged. However, the amplitude of the negative energy mode losing the energy 
grows exponentially. Conversely, the amplitude of the positive energy mode grows because it 
receives energy.
A related example of the mode coupling in the two-fluid dynamics of perturbations has been studied recently by 
\citet{zhuravlev-2019}, who showed that it takes place on the background of the dust settling to the midplane of a
protoplanetary disc.

\subsection{Changing of resonance type}
\label{sec_res_type}

\begin{table}
\caption{Map of the analytical limiting cases for the resonance of modes. The parameter $\tau$ 
increases from the left to the right column, while it is assumed that $\tau\ll 1$. 
It is implied that $\tau$ remains constant for each column, while 
$\hat V_{||}$ increases from top to the bottom of the columns. The upper index after ${\cal R}_{2,3}$ denotes the 
number of the limiting case collected in Sections \ref{sec_tr_res} and \ref{sec_d_res}.}
\label{tab_1}
\begin{tabular}{lllll}

$f < \tau \ll \tau^\prime$ & $\tau \sim \tau^\prime$ & $\tau^\prime \ll \tau$\\

\hline

$0 \leq \hat V_{||}\ll \tau:  {\cal R}_3^{\rm i}$ & $0 \leq \hat V_{||}\ll \tau:  {\cal R}_3^{\rm i}$ & 
$0 \leq \hat V_{||}\ll \hat V_{||}^{\prime}:  {\cal R}_3^{\rm i}$ \\
$\tau \ll \hat V_{||} \ll \hat V_{||}^{\prime\prime}: {\cal R}_3^{\rm ii}$ & $\tau \ll \hat V_{||} \ll 1: {\cal R}_2^{\rm ii} $ & 
$\hat V_{||}^{\prime} \ll \hat V_{||} \ll \tau: {\cal R}_2^{\rm i}$ \\
$\hat V_{||}^{\prime\prime} \ll \hat V_{||} \ll 1: {\cal R}_2^{\rm ii} $ & $\hat V_{||} \to 1: {\cal R}_2^{\rm iii} $ & 
$\tau \ll \hat V_{||} \ll 1:  {\cal R}_2^{\rm ii}$\\
$\hat V_{||} \to 1: {\cal R}_2^{\rm iii}$ & ----------- & $\hat V_{||}\to 1:  {\cal R}_2^{\rm iii}$\\

\hline

\end{tabular}
\end{table}

\begin{table}
\caption{The same as in the Table \ref{tab_1} for $\tau\gtrsim 1$.}
\label{tab_2}
\begin{tabular}{lllll}

$\tau \sim 1$ & $1 \ll \tau < f^{-1}$ \\

\hline

$0 \leq \hat V_{||}\ll \hat V_{||}^{\prime}:  {\cal R}_3^{\rm i}$ & 
$0 \leq \hat V_{||}\ll \hat V_{||}^{\prime}:  {\cal R}_3^{\rm i}$ \\

$\hat V_{||}^{\prime} \ll \hat V_{||}\ll 1:  {\cal R}_2^{\rm i}$ & 
$\hat V_{||}^{\prime} \ll \hat V_{||}\ll 1:  {\cal R}_2^{\rm i}$ \\

$\hat V_{||} \to 1: {\cal R}_2^{\rm iii}$ & 
$\hat V_{||} \to 1: {\cal R}_2^{\rm iv}$ \\

------------ & 
$\hat V_{||} \to 1: {\cal R}_2^{\rm iii}$ \\

\hline

\end{tabular}
\end{table}

\begin{figure}%{h}
\begin{center}
\includegraphics[width=9cm,angle=0]{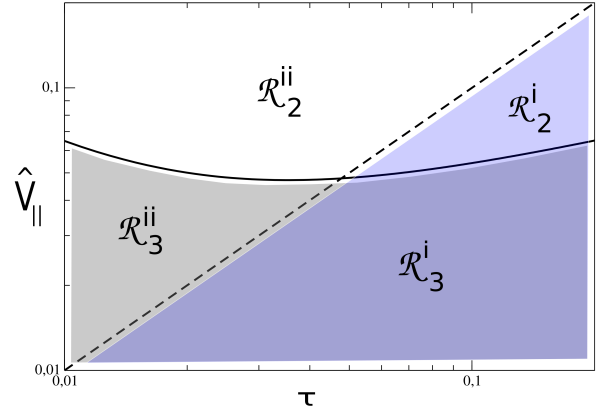}
\end{center}
\caption{Graphic representation of the limiting cases for the resonance of modes for $\hat V_{||}\ll 1$. 
The solid line is given by equation (\ref{V_cr}) for $f=0.01$, while the dashed line shows $\hat V_{||} = \tau$.}
\label{fig_6}
\end{figure}

The restrictions on the existence of ${\cal R}_3$ and ${\cal R}_2$, see equations 
(\ref{Delta_gg_om}) and (\ref{Delta_ll_om}), lead to the
condition for the change of the resonance type 
\begin{equation}
\label{gen_cond_f_cr}
2\omega_c \sim|\Delta|, 
\end{equation}
which yields the critical value of the dust fraction
\begin{equation}
\label{f_cr}
f^\prime = \frac{8\,\omega_c^3}{ \omega_c^2 + \tau^2}.
\end{equation}

It is instructive to rewrite equation (\ref{f_cr}) in terms of the drift velocity. 
One obtains in the leading order over the small dust fraction, $f\ll 1$:
\begin{equation}
\label{om_c_cr}
\omega^\prime_c \approx \frac{1}{2}{f}^{1/3} \tau^{4/3} + \frac{f}{24},
\end{equation}
where $\omega^\prime_c$ denotes the critical value of the mode crossing frequency corresponding to transition
between ${\cal R}_2$ and ${\cal R}_3$ for specified $f$ and $\tau$.
The corresponding critical drift velocity reads
\begin{equation}
\label{V_cr}
\hat V_{||}^{cr} \approx \frac{1}{2} {f}^{1/3} \tau^{1/3} + \frac{f}{24\tau}
\end{equation}
provided that
\begin{equation}
\label{tau_range}
f^{-1} \gg \tau \gtrsim f.
\end{equation}
The latter condition guaranteers that (\ref{V_cr}) remains small compared to unity. Otherwise, equation (\ref{V_cr})
is not valid anymore. As soon as $\hat V_{||} \ll \hat V_{||}^{cr}$, the modes come into ${\cal R}_3$, whereas in the 
other case they come into ${\cal R}_2$.
Equation (\ref{V_cr}) contains two terms, which are denoted as
\begin{equation}
\label{V_cr_p}
\hat V_{||}^\prime \equiv \frac{1}{2} f^{1/3} \tau^{1/3}
\end{equation}
and
\begin{equation}
\label{V_cr_pp}
\hat V_{||}^{\prime\prime} \equiv \frac{f}{24\tau}.
\end{equation}

As long as $\tau \sim \tau^\prime$, where
\begin{equation}
\label{tau_cr}
\tau^\prime \equiv (f/8)^{1/2},
\end{equation}
$\hat V_{||}^\prime \sim \hat V_{||}^{\prime\prime} \sim \tau$.
In the limit of small $\tau \ll \tau^\prime$ there is an inequality 
$\hat V_{||}^{\prime\prime} \gg \hat V_{||}^{\prime} \gg \tau$, 
whereas in the limit of large $\tau \gg \tau^\prime$, conversely, 
$\hat V_{||}^{\prime\prime} \ll \hat V_{||}^{\prime} \ll \tau$.
Thus, the drift velocity corresponding to the transition ${\cal R}_2 \leftrightarrow {\cal R}_3$ approximately equals 
$\hat V_{||}^\prime$ and $\hat V_{||}^{\prime\prime}$ in the case when $\tau$ is, 
respectively, smaller and larger that $\tau^\prime$.
Consequently, as the dust drift intensifies in the case $\tau \gg \tau^\prime$,  the transition from 
${\cal R}_3$ to ${\cal R}_2$ occurs in the strong self-gravity regime, $\hat V_{||} \ll \tau$, at the drift velocity 
given by equation (\ref{V_cr_p}). The corresponding approximate expressions for $\Delta$ 
given by equations (\ref{tr_res_simple}) and (\ref{tau_V_1}) continuously replace one another at 
$\hat V_{||} = \hat V_{||}^\prime$. On the other hand, 
as the dust drift intensifies in the case $\tau \ll \tau^\prime$,  the transition from 
${\cal R}_3$ to ${\cal R}_2$ occurs in the weak self-gravity regime, $\hat V_{||} \gg \tau$, at the critical drift velocity 
given by equation (\ref{V_cr_pp}). Again, the corresponding approximate expressions for $\Delta$ 
given by equations (\ref{tr_res_simple_weak}) and (\ref{tau_V_2}) continuously replace one another at 
$\hat V_{||} = \hat V_{||}^{\prime\prime}$

For a clear exposition of the analytical results, the overall picture of modal resonances for any $\tau$ and $\hat V_{||}$
is shown in the Tables \ref{tab_1}, \ref{tab_2} and graphically in Figure \ref{fig_6}.

It is worth comparing the lower estimate of the growth rate in ${\cal R}_2$, see 
equation (\ref{min_growth_double_res}) (or equation (\ref{coupl_Delta})), with the approximate 
growth rate in ${\cal R}_3$ in the absence of the dust drift, $\hat V_{||}=0$, see equation (\ref{tr_res_simple}). 
They are comparable to each other provided that $\tau \sim  f^{1/2}$, which is of the order of $\tau^\prime$.
Hence, in the limit of $\tau \gg \tau^{\prime}$ ${\cal R}_3$ taking place at $\hat V \ll \hat V_{||}^\prime$ 
provides a higher growth rate than ${\cal R}_2$ taking place, at least, for $\hat V \sim \tau \gg \hat V_{||}^{\prime}$, i.e. 
sufficiently close to the mode coupling, see Section \ref{sec_coupl}. This trend is increasing as $\tau$ approaches unity
or becomes higher than unity.

\subsection{Connection with \citetalias{squire-2018_acoustic} instability}
\label{sec_compl}

Let the general dispersion equation (\ref{disp}) be considered in the absence of self-gravity, $\tau=0$, 
and, additionally, in the limit of the small dust fraction. 
Once $\Delta$ is a small deviation from an exact solution of equation (\ref{disp}) taken in the limit $f\to 0$ as 
$\omega = \hat V_{||} k$, it obeys the following equation 
\begin{equation}
\label{Delta_SH_eq}
(\hat V_{||} k + k + \Delta) (\hat V_{||} k - k + \Delta) \Delta = {\rm i} f \hat V_{||}^2 k^2
\end{equation}
derived in the leading order in $f$. The restriction 
\begin{equation}
\label{restr_SH_3}
\Delta \gg k
\end{equation}
yields the reduced equation for $\Delta$
\begin{equation}
\label{eq_R_3}
\Delta^3 = {\rm i} f \hat V_{||}^2 k^2, 
\end{equation}
and the corresponding solution
\begin{equation}
\label{sol_SH_R_3}
\Delta = \left \{ - {\rm i},\, \frac{\pm \sqrt{3} + {\rm i} }{2} \right \} f^{1/3} k^{2/3} \hat V_{||}^{2/3}.
\end{equation}
It is seen that equation (\ref{sol_SH_R_3}) recovers equation (18) of \citetalias{squire-2018_acoustic}. 
Therefore, the long-wavelength \citetalias{squire-2018_acoustic} instability can be considered as
${\cal R}_3$ in the non-self-gravitating medium.
The solution (\ref{sol_SH_R_3}) is also identical to equation (\ref{tr_res_simple_weak}) 
after the replacement $k \to \tau$. This is because $\tau$ standing in equation (\ref{tr_res_simple_weak}) originates from 
$k_c \approx \tau >0$ in the limit of the small drift velocity. 
However, since $\omega_s(k_c) \approx 0$ in the self-gravitating medium, 
the corresponding restriction on the occurrence of ${\cal R}_3$ is much weaker for the subsonic drift. 
Indeed, the validity of equation (\ref{eq_R_3}) follows from the restriction (\ref{restr_SH_3}) combined with the 
solution (\ref{sol_SH_R_3}). Explicitly,
\begin{equation}
\label{restr_SH_3_expl}
k \ll f \hat V_{||}^2.
\end{equation}
At the same time, equation (\ref{tr_res_simple_weak}) is valid as far as $\hat V_{||} \ll \hat V_{||}^{\prime\prime} \ll 1$, 
which reads 
\begin{equation}
\label{V_cr_R_3_new}
k_c \ll \frac{f}{24 \hat V_{||}}
\end{equation} 
after making use of equation (\ref{V_cr_pp}).
Thus, the weak self-gravity of gas-dust medium with the subsonic bulk drift of the dust extends the resonance of three modes
and the corresponding value of the growth rate onto much shorter length-scales up to $k_c \sim \tau^{\prime}$, see
equation (\ref{tau_cr}).
%Поэтому, при увеличении скорости дрейфа - SH неуст в режиме R_3 подходит (воспроизводит) к воображаемой линии моего R_3
%уже тогда, когда мои инкременты при той же новой скорости - уходят на режим R_2. 

As soon as $\hat V_{||} \to 1$, equation (\ref{Delta_SH_eq}) reduces to
\begin{equation}
\label{eq_R_2}
\Delta^2 = {\rm i} f \frac{k}{2}
\end{equation}
provided that 
\begin{equation}
\label{restr_SH_2}
(1-\hat V_{||}) k \ll \Delta \ll k.
\end{equation}
The solution of equation (\ref{eq_R_2}) reads
\begin{equation}
\label{sol_SH_R_2}
\Delta = \pm \frac{1 + {\rm i}}{2} f^{1/2} k^{1/2},
\end{equation}
which recovers the dependence $\propto k^{1/2}$ of the supersonic acoustic RDI at intermediate wavelengths 
shown in Figure 1 of \citetalias{squire-2018_acoustic}. 
Therefore, the intermediate-wavelength \citetalias{squire-2018_acoustic} instability in the transonic regime 
can be considered as ${\cal R}_2$ in the non-self-gravitating medium. Since there is no mode crossing in the absence
of self-gravity, the sound wave (SW hereafter) and SDW propagating in the same direction 
can come into resonance provided that the dust fraction 
is sufficiently high, or conversely, the drift velocity is sufficiently close to the sonic value,
which is expressed by the LHS of inequality (\ref{restr_SH_2}).   
The solution (\ref{sol_SH_R_2}) is also identical to equation (\ref{tau_less_1}) 
after the replacement $k \to k_c \approx \tau/(2\delta)^{1/2}$. Note that because this solution
is produced by the mode crossing of HSW and SDW, it is valid in the regime of weak self-gravity
for any non-zero dust fraction. By the same reason, the counterpart of this solution in the subsonic regime 
given by equation (\ref{tau_V_2}) is also valid for any small dust fraction as well as for any small
$\hat V_{||} \gg \max \{ \tau, \hat V_{||}^{\prime\prime} \}$: see the tables \ref{tab_1} and \ref{tab_2}.

An explicit form of the restriction (\ref{restr_SH_2}) after it is combined 
with the solution (\ref{sol_SH_R_2}) reads
\begin{equation}
\label{restr_SH_2_expl}
k \gg f \gg (1-\hat V_{||}) k. 
\end{equation}
The RHS of this inequality implies that ${\cal R}_2$ in the non-self-gravitating medium recovers the growth rate 
attained in the corresponding regime of weak self-gravity, i.e. both ${\cal R}_2^{\rm ii}$ and ${\cal R}_2^{\rm iii}$, 
see the tables \ref{tab_1} and \ref{tab_2}, for $\hat V_{||}$ very close to sonic value only. 
The latter explains the ${\cal R}_2$ bumps on the profiles of the
growth rate for $\tau=0.01, 0.1, 1.0$ substantially exceeding 
the profile of \citetalias{squire-2018_acoustic} instability in panels (c-f) in Figure \ref{fig_1}.
In other words, while the drift velocity gradually increases up to the sonic value, the crossing modes approach each other
farther and farther from $k=k_c$ causing the expansion of the zone of resonance around $k=k_c$, which
is seen as the growth rate of \citetalias{squire-2018_acoustic} instability approaching the bumps from below.

%Но этот квадратный корень начинается ``гораздо раньше'': уже в формуле \ref{tau_V_2}, при V<<1. 
%Это уже работает R_2. А без гравитации - у HS-18 - без mode crossing, инкремент быстро падает при уменьшении V, 
%см. рисунки Im_...

% !!! Поскольку условие triple resonance для неуст. HS-18 - это k < f V^2, то для меня в резонансе это означает \tau < f,  
%     т.е. выход за пределы приближения V^\prime\prime << 1. Поэтому, все что садится у меня на сплошную кривую, а это  
%     \tau > f - садится уже в режиме double coupling !!! 

% Общее свойство по поводу подхода кривой HS-18 к моим оценкам в резонансах: и для triple и для double coupling 
% я выхожу на них (это мой режим weak self-gravity и для R_3 и для R_2) гораздо раньше по параметрам, чем они --- поскольку 
% мне помогает сдвиг mode crossing на k = k_J (у них эта точка всегда в k=0) !!!

% Но оценки инкремента действительно одни и те же с заменой \tau на k (для рассм. мной случая V<<1 только).

\section{Astrophysical implications}
\label{sec_physics}

The Epstein aerodynamic drag yields the stopping time \citep{weidenschilling-1977}:
\begin{equation}
\label{Eps_drag}
t_s = \frac{\rho_m s}{\rho_g v_{th}},
\end{equation}
where $s$ is the grain size and the mean thermal velocity $v_{th} = (8/\pi)^{1/2} c_s$.

Estimate of $\tau$ with the help of equation (\ref{Eps_drag}) reads
\begin{equation}
\label{estimate_tau}
\tau \simeq 0.25 \left ( \frac{s}{10^{-4}{\rm cm}} \right ) \left ( \frac{\rho_m}{3\,{\rm g\, cm}^{-3}} \right ) 
\left ( \frac{T}{50 {\rm K}} \right )^{-1/2} \left ( \frac{n}{50\,{\rm cm}^{-3}} \right )^{-1/2},
\end{equation}
where the temperature and gas density are normalised by plausible average values 
for the young molecular cloud a few 
Myr old, when its Jeans mass falls below its own mass, see e.g. numerical simulation by \citet{vazquez-semadeni-2007}.
In the course of its evolution, a molecular cloud fragments into denser and cooler structures. It finally gives birth
to prestellar cores with typical density and temperature $n \simeq 10^{5} {\rm cm}^{-3}$ and $T\simeq 10{\rm K}$, respectively, see e.g. numerical simulation by \citet{inutsuka-2000}. Therefore, in prestellar cores $\tau \simeq 0.01$
indicates that the micron-sized grains are stronger coupled to the gas as the dynamical timescale is defined 
by the self-gravity of the gas-dust mixture.   
% $\tau > 1$ можно натянуть только для больших частиц, т.е. если в ISM предварительно росла пыль (возможно или нет?). 
The range $0.01 \lesssim \tau \lesssim 0.25$ may shift both up and down depending on the actual grain size.
According to the conventional dust model in the diffuse ISM, see \citet{draine-2003}, 
the peak of the grain size distribution is attained at sub-micron scales, $s\simeq 0.3 \mu m$.
The size of grains in clouds with $n > 100\, {\rm cm}^{-3}$ is less clear, as they are able to accrete volatile
elements increasing both $s$ and $\tau$ and consequently the mass fraction accumulated 
in dust \citep{kohler-2015}. Moreover, in the dense regions of molecular clouds (cores), 
where $n \sim 10^4\,{\rm cm}^{-3}$ and higher,
grains coagulate with each other attaining sizes as large as $s \sim 10\mu m$, see e.g. \citet{ormel-2009}.
This is confirmed by the observational evidence for micron-sized grains in molecular clouds, see e.g. 
\citet{pagani-2010}, \citet{saajasto-2018}, as well as by the indication of the dust fraction variations 
within a particular molecular cloud \citep{liseau-2015}. The following conclusions about the growth rate of 
the new resonant instabilities are made for nominal $\rho_m = 3\,{\rm g\, cm}^{-3}$, $s = 1 \mu m$ and 
$f =0.01$ though various $\tau$ and $\hat V_{||}$ with the help of the tables \ref{tab_1} and \ref{tab_2}.

As long as $\tau \lesssim \tau^\prime \simeq 0.035$ for $f=0.01$, see equation (\ref{tau_cr}), 
the lower estimate of the growth rate (given in units of the inverse free fall time in this 
Section) is $\sim 0.22 \tau^{1/3}$ attained 
in the regime ${\cal R}_3^{\rm i}$ for the sufficiently slow dust drift  $\hat V_{||} \lesssim \tau$ . 
This value weakly depends on the grain size and yields the lowest value of the growth rate $\sim 0.05$ for $\tau=0.01$. 
Further, the largest growth rate attained in the regime ${\cal R}_3^{\rm ii}$ for $\hat V_{||}\simeq 0.04$ 
corresponding to the change of the resonance type in this case (see equation (\ref{V_cr_pp}) )
is equal to $\sim 0.12$. For a higher drift velocity, e.g. $\hat V_{||} = 0.1$, one should use the estimate 
of the growth rate for the regime ${\cal R}_2^{\rm ii}$ obtaining $\sim 0.32$, which is almost two orders of 
magnitude higher than the corresponding growth rate of the intermediate wavelength \citetalias{squire-2018_acoustic} 
instability, see panel (b) in Figure \ref{fig_1}. 
As discussed in Section \ref{sec_compl}, equations (\ref{tr_res_simple_weak}) and
(\ref{tau_V_2}) recover the corresponding results of \citetalias{squire-2018_acoustic} obtained for the non-self-gravitating
medium. However, they are valid for the deeply subsonic dust drift, since SDW falls in resonance with HSW, which
is (are) slowed down by the self-gravity of gas. 
As a less coupled dust and gas are considered, $\tau \gtrsim 0.035$ for $f=0.01$, the lower estimate 
of the growth rate is simply the square root of the dust fraction, $\sim 0.1$, given by equation
(\ref{min_growth_double_res}). This value is attained at $\hat V_{||} \sim \tau$ and independent of the grain size.
For example, as $\hat V_{||}=0.5$, the growth rate is an order of magnitude higher than that of 
intermediate-wavelength \citetalias{squire-2018_acoustic} instability, see panel (c) in Figure \ref{fig_1}.
If there is no dust drift, $\hat V_{||} \to 0$, the growth rate becomes $\sim 0.14$ for a particular value $\tau=0.25$. 
On the other hand, as $\hat V_{||} \to 1$ the growth rate for $\tau=0.25$ 
behaves as $\sim 0.1 (2\delta)^{-1/4}$, where $\delta \equiv 1-\hat V_{||}$, see equation (\ref{tau_less_1}).

Thereby, in all cases discussed just above, the growth rate of the new instability is 
a considerable fraction of $t_{ff}^{-1}$. Also, it weakly depends on the drift velocity
covering the whole subsonic band from $0$ to $1$.
As the size of the cloud passes the Jeans length in the course of its formation, its gravitational contraction 
proceeds on the longer time-scale as compared with $t_{ff}$. The resonant instability due to either 
${\cal R}_2$ or ${\cal R}_3$ may have enough time to launch the collapse of the sub-Jeans mass cloud. 
How  much the new instability affects the gravitational collapse of the cloud, will depend 
on its cooling rate.

It is important to note that the new instabilities provide the growth of the dust fraction. Yet, it should be
emphasised that the gas is also considerably affected by the instability in spite of the small value 
of the background dust fraction. For modes, this means that the amplitude of the relative perturbation of 
the gas density is not
negligible in comparison with that of the relative perturbation of the dust fraction.
The latter can be checked using equation (\ref{eq_delta}), which is taken for the growing modes 
at ${\cal R}_3$ and ${\cal R}_2$. It is found that the ratio of the Fourier amplitudes 
$\bar \delta$ and $\bar h^\prime$ is finite provided that both $f$ and $\tau$ (or $\hat V_{||}$) are finite. 
For example, in the case $\hat V_{||}=0$ and $\tau \ll 1$ it reads
\begin{equation}
\label{fourier_delta_h}
\bar \delta \approx \frac{\tau^{2/3}}{f^{1/3}} \bar h^\prime.
\end{equation}
In the limit $f\to 0$ the inequality $\bar \delta \gg \bar h^\prime$ is always satisfied, so that the instability 
of the gas-dust mixture is provided mostly by the relative perturbation of the dust density, $\rho_p^\prime/\rho_p$, 
while that quantity for the gas is negligible. However, the values of $f$ and $\tau$ present in the ISM,
as well as their low powers entering equation (\ref{fourier_delta_h}), make $\delta$ and $h^\prime$ 
comparable to each other, especially in dense clouds.
%But as far as the Jeans scale is of the order of the size of collapsing domain, the collapse is not a free fall 
%and the new instability operates promoting the dust overdensity. 
As the Jeans scale becomes much smaller than the size of the cloud, the gravitational contraction 
enters the free fall stage, when the equivalence principle of gravity freezes the grains into common free
motion of matter, i.e. the growth of the dust fraction is stalled, see the comments in the Appendix \ref{sec_lw}.
How strong the concentration of dust gets during the time passed from the sub-Jeans contraction of the cloud up to its free 
fall is an issue for future research.
As soon as the bulk drift of the dust is significant, the growing dust overdensities are carried by the unstable 
gas-dust wave which is akin to SDW. During the non-linear stage occurring in a sufficiently dense cloud, 
these dust overdensities may become opaque to ambient radiation forcing the dust drift. 
At this moment, the dust drift is suppressed and the dust overdensities lag behind the wave front, 
which means that the resonance ceases to operate. The gas-dust wave becomes an ordinary HSW running away from
dusty domains. The latter effect additionally increases the final dust fraction as it removes dense gas out of 
the domains with the enhanced dust concentration. 
This is an other way that dusty domains may be produced at the sub-Jeans scales 
of dense clouds.

% discuss primordial discs now - what can be explained by the overdensities of dust in molecular clouds?
%Recent ALMA observations of rather big primordial discs, age $5\times 10^4$yrs...

The dusty prestellar cores, which may form within the suggested scenarios, may potentially give birth to big 
young protoplanetary discs (see \citet{lee-2017} and \citet{lee-2018} for recent observations of such systems). 
Though the majority of these Class 0 discs have smaller sizes, see \citet{ansdell-2018}, the formation
of bigger discs is not exceptional. Moreover, it is challenging from a theoretical view, see \citet{hennebelle-2016}
and the review by \citet{zhao-2020}. 
Numerical simulations of the magnetised core collapse have shown that the magnetic braking 
should prevent the young disc formation when dissipation of the magnetic flux is neglected, see \citet{mellon-2008}.
Considering the dissipative effects corrects the situation (see e.g. \citet{masson-2016}\footnote{For the most recent
studies see also  \citet{machida-2019} and \citet{lam-2019}}). However, it 
reveals that Class 0 disc size is defined mostly by the ambipolar diffusion of magnetic field, which in turn
considerably depends on the distribution of the grains' size (see \citet{zhao-2016}, \citet{dzyurkevitch-2017} and \citet{zhao-2018}). The amount of very small charged grains may be reduced by the dust coagulation, which 
occurs more efficiently in the dusty environment \citep{ormel-2009}.

\section{Summary}

A partially coupled gas-dust mixture is subject to GI at scales smaller than the Jeans 
length-scale.
An unbounded uniform medium becomes significantly unstable for a small fraction of dust, which can be
explained by the resonant nature of the new instability. The scale of resonance is defined by 
the crossing of gas modes (two oppositely propagating HSW) with the dust mode (SDW) 
existing in a mixture in the limit of the negligible dust fraction, $f\to 0$.
As the dust is suspended in gas, SDW is formally a static wave coming to resonance with both HSW.
This resonance denoted here as ${\cal R}_3$
occurs strictly at the Jeans scale, $k_c=k_J$, see Section \ref{sec_tr_res}. 
The growth rate of GI corresponding to ${\cal R}_3$ evaluated 
as the solution of the reduced dispersion equation (\ref{res_disp}) is determined by $f^{1/3}$. 
As there is subsonic bulk drift of the dust, $0< \hat V < 1$, SDW propagates with phase velocity 
equal to $\hat {\bf V}$ projected onto wavevector denoted as $\hat V_{||}$. 
In this case, it comes into resonance with HSW propagating in the same 
direction with equal phase velocity. This resonance denoted here as ${\cal R}_2$ takes place at $k_c>k_J$, 
see Section \ref{sec_d_res}. According to equation (\ref{res_disp}), the growth rate of GI corresponding to ${\cal R}_2$ 
is determined by $f^{1/2}$. 

An important free parameter of the considered model is the ratio of the stopping time of the grains 
to the free fall time, $\tau$, 
which quantifies the relative strength of the dust to gas dynamical coupling and the mixture self-gravity.
The analytical approach to the problem shows that simple estimates of the growth rate can be obtained in the 
regimes of weak self-gravity, $\tau \ll \hat V \ll 1$, and strong self-gravity, $\hat V_{||} \ll \tau$ along with
$\hat V_{||} \ll 1$, see equations (\ref{tr_res_simple}-\ref{tr_res_simple_weak}) and 
(\ref{tau_V_1}-\ref{tau_V_2}) for the resonances ${\cal R}_3$ and ${\cal R}_2$, respectively. 
These estimates may be useful in various applications. The general picture of all limiting cases
is represented in tables \ref{tab_1} and \ref{tab_2} and, additionally, in Figure \ref{fig_6}. 
For particular values of $\tau$ and $f$, there is a critical value 
of the drift velocity, $\hat V_{||}^{cr}$, which corresponds to the boundary between ${\cal R}_3$ and ${\cal R}_2$, 
see equation (\ref{V_cr}). The critical drift velocity takes the minimum $\hat V_{||}^{cr} \sim \tau$ at 
$\tau \sim \tau^\prime$ given by equation (\ref{tau_cr}). The analytical results show that for a particular, sufficiently
small $\tau \lesssim \tau^\prime$, the lower estimate of the growth rate as function of the subsonic drift velocity 
equals to $f^{1/3} \tau^{1/3} t_{ff}^{-1}$. This value is attained for ${\hat V_{||} \lesssim \tau}$.  
Otherwise, as $\tau \gtrsim \tau^\prime$, such a lower estimate
equals to $f^{1/2} t_{ff}^{-1}$, which is attained at ${\hat V_{||} \sim \tau}$.
It is discussed that in molecular clouds the growth rate of GI of the gas-dust mixture at sub-Jeans scales can attain significant fraction of the inverse free fall time, see Section \ref{sec_physics}.

The growth rate of GI in the vicinity of resonance is determined by the coupling term given by equation (\ref{eps}).
For small $f \ll 1$, the coupling term reduces to equation (\ref{eps_res}) at the mode crossing, 
which is used to obtain the approximate growth rates for ${\cal R}_3$ and ${\cal R}_2$, see equations 
(\ref{Delta_triple}) and (\ref{Delta_double}), respectively. 
As one follows the origin of the coupling term from the general equations (\ref{eq_h_1}-\ref{eq_v}), it becomes
clear that the coupling term emerges from the gravitational dust back-reaction
on gas in the regime of strong self-gravity. The latter is introduced by the additional small gravitational attraction 
provided by an excess of the dust in gas-dust mixture. In turn, an excess of dust is caused by the relative velocity
of dust to gas in the perturbed dynamics. Therefore, in the regime of strong self-gravity (including the case 
$\hat V_{||}=0$), dust back-reaction on the gas has nothing to do with the bulk drift of the dust. 
On the other hand, the coupling term emerges from the aerodynamical dust back-reaction on gas in the regime of weak self-gravity. The latter is introduced by perturbations of the density of dust moving at the velocity of the bulk drift, which is 
similar to the case of the \citetalias{squire-2018_acoustic} instability. Moreover, the corresponding estimates 
of the growth rates in ${\cal R}_3$ and ${\cal R}_2$ recover that of \citetalias{squire-2018_acoustic} after the
replacement $k \to k_c$ given by equation (\ref{k_c}). Indeed, the long- and intermediate-wavelength \citetalias{squire-2018_acoustic} instability may be considered as, respectively, ${\cal R}_3$ and ${\cal R}_2$
in the non-self-gravitating medium, see Section \ref{sec_compl}. However, it becomes much stronger on account of 
self-gravity due to the mode crossing of HSW and SDW at $k_c > k_J$. In contrast, the ordinary SW always have 
higher phase velocity than SDW\footnote{Technically, in this case the mode crossing is located at $k=0$} 
requiring the much higher dust fractions for these mode to come into resonance. 
Additional remarks on the physics of the long-wavelength \citetalias{squire-2018_acoustic} are given in Appendix 
\ref{sec_HS18}.

% О частном случае спаривания мод.
% Собственно про механизм неустойчивости по ссылкой на энергию возмущений при mode coupling.

Generally, the coupling term (\ref{eps_res}) takes a complex value, which causes the avoided crossing of modes
akin to HSW and SDW for $f>0$, see Section \ref{sec_stream_dust} and the top-right panel 
in Figure \ref{fig_2} as well as the top panels in Figures \ref{fig_3} and \ref{fig_5}.
However, for the particular value of the drift velocity given by equation (\ref{coupl_cond}), the 
coupling term (\ref{eps_res}) becomes real, which means that as $f>0$, the crossing HSW and SDW give birth 
to the coupled modes represented by a complex conjugate pair of solutions of the reduced dispersion equation,
see Section \ref{sec_coupl}.
It is not difficult to gain further insight into the physics of GI, which appears as bumps on curves 
in the bottom panels in Figure \ref{fig_4} in this case. The energy of modes akin to 
HSW and SDW outside the band of the mode coupling can be obtained assuming that they are approximately neutral, 
see the Appendix \ref{sec_mode_energy} for the description of the corresponding simplified model. The modes 
coalescing inside the band of GI have the energies of the opposite signs. Thus, the conserved total 
energy flows from the negative energy mode (SDW) to the positive energy mode (HSW) 
causing the growth of their amplitudes.

GI of the gas loaded with a suspended dust which is produced by ${\cal R}_3$ for strictly $\hat V_{||}=0$
can be considered as the missing link in the description of the 
dynamics of a partially coupled self-gravitating gas-dust mixture. This is the link between the limiting case 
of the free falling dust-laden gas corresponding to $k \to 0$, see the Appendix \ref{sec_lw}, and 
the limiting case of the dust settling through the gas being in hydrostatic equilibrium in its own gravitational well 
corresponding to $k\to \infty$, see the Appendix \ref{sec_sw}. 
This is illustrated by the dot-dashed curve of the GI growth rate in the bottom-left panel in Figure \ref{fig_2}.
%...a general view of GI of gas-dust mixture is given in Figures \ref{fig_1}-\ref{fig_3} showing the accurate solution
%of the full dispersion equation (\ref{disp}).... 

% В конце - о перспективах направления

In future, the dynamic role of dust in the GI of dense ISM should be studied in the framework of 
the global linear stability analysis of real configurations. This will help to see, whether the self-gravitating gas-dust mixture with a small fraction of dust can be most unstable with respect to perturbations growing due to 
${\cal R}_3$ or ${\cal R}_2$. Of course, accompanying numerical simulations are required.

GI of a mixture with suspended dust, $\hat V_{||}=0$, considered at the Jeans scale induces the 
growing relative velocity of dust and gas, ${\bf v}$. As the grain size is distributed in some range, the 
growing relative velocity of grains of different size should trigger the enhanced growth of grains via 
coagulation. In turn, the growth of grains should cause the increase of their stopping time, $t_s$, and consequently, 
the further decoupling of the gas-dust mixture along with the following speedup of its gravitational contraction.
Therefore, the model for dynamics of self-gravitating gas-dust mixture should be extended to account for grains' growth.

On the other hand, turbulence generally weakens dynamic instability. The ISM is also 
turbulent. Its role in damping of the growth rates obtained in this study is still to be determined.

%%%%%%%%%%%%%%%%

The case $\tau>1$ describes weakly coupled dust and gas.
While not relevant for typical conditions in ISM (see Section \ref{sec_physics}), it, however, 
may be appropriate in the densest prestellar cores as nurseries of the big particles $\sim 1\, {\rm mm}$ 
born due to effective coagulation of typical micron-sized grains.
The weak coupling of big particles to the gas invalidates the description of dust as a pressureless fluid. 
For consistency, one should additionally account for the non-zero velocity dispersion of the dust in this case.

%%%%%%%%%%%%%%%%%%%%%%%%%%

The model of gas-dust perturbations considered here does not take into account perturbation of $t_s$ due to 
perturbation of gas density. It can be checked that this extension of the model does not affect the results 
obtained for ${\cal R}_{2,3}$. However, it may be important far from the mode crossing along with the other 
non-resonant corrections contained in the general dispersion equation (\ref{disp}). The non-resonant contribution
to GI of the gas-dust mixture may be important as the dust fraction is not small. This issue can
be addressed in a future work. 

At last, the charged grains are affected by Coulomb drag and Lorentz force if 
mixed with ionised and magnetised gas. GI of such a mixture consisting of the charged dust and weakly ionised plasma 
is another problem to be resolved.

% в эту же копилку:

%
%It is very important, that self-gravity of medium, in some sense, widens the range of RDI found by
%\citetalias{squire-2018_acoustic} to the subsonic drift of dust!!! 
%
% Прямо так и начать: что есть новые резонансные неустойчивости в диске - однако, в отсутствие вращения, в гидродинамич. 
% баротропном (т.е. базовом простом общем ) случае, существует только звук, который не будет давать пересечения мод, см. 
% сразу иллюстрацию. Только звуковая скорость дрейфа, см. SH-17, но мол вообще-то это довольно специальные условия (обосновать!).
%
% Но! Звук можно замедлить самогравитацией! Т.е. - рассмотрим задачу Джинса с дрейфом пыли. См. иллюстрацию пересечения мод. 
% Т.е. график Re [\omega] (k) для случая f=0.
%
%First, the effect of the dust drift is considered in perturbations only, letting the background be at rest.
%After that, the stability of mixture is studied in the case when there is bulk drift of the dust under the external forcing, 
%which is physically the ambient starlight pressure...

%%%%%%%%%%%%%%%%%%%%%%%%%%%%%%%%%%%%%%%%%%%%%%%%%%%%%%%%%%%%%%%%%%%%

%это уже и так сказано в аппендиксе?..

% По крайней мере в случае неуст. без дрейфа (triple coupling), для нее необходима гетерогенность смеси, т.е. 
% присутствие эксп. растущей доли пыли в возмущениях. Т.е. постоянное опережение пылинками движения вещества в коллапсе.

%%%%%%%%%%%%%%%%%%%%%%%%%%%%%%%%%%%%%%%%%%%%%%%%%%%%%%%%%%%%%%%%%%%%%%%%

\section*{Data availability}
No new data were generated or analysed in support of this research.

\section*{Acknowledgments}
I thank Natalia Dzyurkevich for discussing the formation of the Class 0 discs.
I am grateful to Artem Tuntsov and Henrik Latter for careful reading of the manuscript and their useful comments and 
suggestions that helped to improve the presentation of the study.
Of course, this work would not have been done if it were not for the dedication of my family.
%The author acknowledges the support from the Program of development of M.V. Lomonosov Moscow State University (Leading Scientific School 'Physics of stars, relativistic objects and galaxies').
I acknowledge the support from the Foundation for the Advancement of Theoretical
Physics and Mathematics ``BASIS''.
Additionally, this work was supported in part by the Government and the Ministry of Science and Higher Education of the Russian Federation (project no. 075-15-2020-780) and in part by the Program of development of Lomonosov Moscow State University.

\bibliography{bibliography}

\appendix 

\section{Towards interpretation of GI of dust-laden medium}
\label{sec_mixt_GI}

It is assumed here that dust with small fraction, $f\ll 1$, is suspended in gas, $\hat V_{||}=0$.
The dynamics of small perturbations of self-gravitating homogeneous medium is considered in the three basic cases, 
$k\to 0$, $k\approx k_J$ and $k\to \infty$. In all these cases, the analysis starts from equations (\ref{eq_h_1}-\ref{eq_v})
and it is assumed for simplicity that 
\begin{equation}
\label{t_ev_s}
t_{ev} \gg t_s,
\end{equation}
where $t_{ev}$ is the characteristic time of evolution of 
a mixture.

%{\bf Check the dimensions everywhere below from equations for divergence of $u_g$ and v!!!}

\subsection{Long-wavelength limit}
\label{sec_lw}

As $k\to 0$, the perturbation of the pressure gradient becomes negligible and the gas 
undergoes free fall uniformly with the dust. 
Clearly, there is no perturbation of the relative velocity of gas and dust, 
which is demonstrated by equation (\ref{eq_v}), where all terms except those containing $\nabla \cdot{\bf v}$ cancel
each other by means of equation (\ref{eq_h_1}).
As $\nabla\cdot {\bf v}\to 0$ drives the relative perturbation of the dust fraction, $\delta$, see equation (\ref{eq_delta}), 
it is clear that
\begin{equation}
\label{delta_ff}
\delta \approx 0
\end{equation} 
in the course of collapse. In this way, one obtains the growth 
rate of gas-dust density approaching 
\begin{equation}
\label{k_0}
\omega \to {\rm i} \tau \left ( 1 +\frac{f}{2} \right )
\end{equation}
in units of $t_s$. This equation recovers the inverse free fall time generally expected for the gas-dust mixture 
with grains frozen in fluid elements of gas. 

This may be the reason for a weak segregation of (sub-)micron-sized grains with gas
found in the numerical simulations of collapse of gravitationally unstable Bonnor-Ebert sphere laden with 
partially coupled dust, see \citet{bate-2017}.

\subsection{Short-wavelength limit}
\label{sec_sw}

As the wavenumber becomes high, $k\to\infty$, while $t_{ev}$ is kept constant, the amplitude of $h_1$ vanishes. 
This reproduces the incompressible gas with the velocity free of divergence, $\nabla \cdot{\bf u}_g \to 0$.
Thus, equation (\ref{eq_v}) yields $\nabla\cdot {\bf v} \to - f\tau \omega_{ff} \delta$ 
and equation (\ref{eq_delta}) takes the form
\begin{equation}
\label{eq_k_high}
\frac{\partial \delta}{\partial t} \approx f \tau\, \omega_{ff} \delta,
\end{equation}
which describes a slow collapse of self-gravitating dust drifting through the gas in hydrostatic equilibrium. 
Such a collapse is not a free fall: it is restrained by the aerodynamic drag. 
The dimensionless growth rate of this kind of dust clumping reads
\begin{equation}
\label{gr_rate_k_high}
\omega \to {\rm i} f \tau^2.
\end{equation}
It can be checked that equation (\ref{gr_rate_k_high}) is in agreement with an accurate curves plotted at the
panel (a) in Figure \ref{fig_1}.

\subsection{Jeans wavelength}
\label{sec_Jw}

% Here, we derive the reduced equation, which describes the dynamics of gas-dust perturbations...

In the limit of the negligible dust fraction, $f\to 0$, an additional simplifying assumption
\begin{equation}
\label{t_ev_ff}
t_{ev} \gg t_{ff}
\end{equation}
is justified at the considered scale $k \simeq k_J$, since one deals with either a slowly propagating sound wave 
or weak Jeans instability, when
\begin{equation}
\label{Jeans_balance}
\nabla h^\prime \approx {\bf g}_g
\end{equation}
with ${\bf g}_g$ being the gravitational acceleration arising from the gas self-gravity.
Since for small $f \ll 1$ the new solution discussed here slightly differs from this basic case, 
it is reasonable to use the restriction (\ref{t_ev_ff}). 
The assumptions (\ref{t_ev_s}) and (\ref{t_ev_ff}) make the inertial terms on the LHS of equation (\ref{eq_v}) small
compared with the leading terms on the RHS of equation (\ref{eq_v}), i.e. the terms remaining there in the limit $f\to 0$.
Thus, there is an approximate balance
\begin{equation}
\label{GI_TVA}
\nabla \cdot{\bf v} \approx -\tau \, \omega_{ff} h_1.
\end{equation}
%Описать физический смысл этого уравнения.
Equation (\ref{GI_TVA}) resembles the balance of terms in the terminal velocity approximation applicable for certain 
problems of gas-dust dynamics in protoplanetary discs, see e.g. \citet{youdin-goodman-2005} and \citet{zhuravlev-2019}.
Further, equation (\ref{GI_TVA}) implies that the term $\sim \nabla \cdot{\bf v}$ on the RHS of equation (\ref{eq_h_1})
along with the addition $\sim f$ in front of $h^\prime$ in the square brackets on the RHS of this equation  
can be omitted as they are of the order of a small correction to the single-fluid GI growth rate 
due to a slight increase 
of the total density with the account of the dust. 
Also, equation (\ref{GI_TVA}) is used to reduce equation (\ref{eq_delta}).

Finally, the reduced set of equations can be expressed as
\begin{equation}
\label{eq_k_J_1}
\frac{\partial {\bf u}_g}{\partial t} = {\bf g}_p,
\end{equation}
\begin{equation}
\label{eq_k_J_2}
\frac{1}{c_s^2} \frac{\partial h^\prime}{\partial t} = -\nabla \cdot {\bf u}_g, 
\end{equation}
\begin{equation}
\label{eq_k_J_3}
\frac{\partial \delta}{\partial t} = \tau\, \omega_{ff} \frac{h^\prime}{c_s^2},
\end{equation}
\begin{equation}
\label{eq_k_J_4}
\nabla \cdot {\bf g}_p = -f\omega_{ff}^2 \delta,
\end{equation}
where ${\bf g}_p \equiv - \nabla \Phi_p$ is the gravitational acceleration caused by deviation of the dust fraction from its background value. 
It is ${\bf g}_p$ standing on the RHS of equation (\ref{eq_k_J_1}) that represents the dust gravitational back-reaction on gas
making the coupling term entering the dispersion equation (\ref{disp}) non-zero in the absence of the 
bulk drift of the dust, see the description of the corresponding resonant instability in Section \ref{sec_tr_res}.

The spatially harmonic solution of equations (\ref{eq_k_J_1}-\ref{eq_k_J_4}) corresponding to exponential growth of perturbations with the growth rate equal to $(f \tau)^{1/3} \omega_{ff}$ reads
\begin{equation}
h^\prime,\, \delta \propto \cos {\bf k \cdot r}_* \quad \mbox{and} \quad u_g,\, g_p \propto - \sin {\bf k \cdot r}_*. 
\end{equation}
 
It is worth noting that $\delta > 0$ means not just the increase of the dust density
as compared to its background value, but rather the increase of an excess of the dust density compared to the gas density, 
$\delta_p > h^\prime/c_s^2$, in the course of collapse. This excess is continuously generated by perturbation 
of the relative velocity of dust with respect to gas, which, in turn, arises due to gravitational attraction 
of grains into the potential well of gas overdensities, see equation (\ref{GI_TVA}). This is the way how an incremental 
gravitational potential, $\Phi_p$, is produced. If it were not for the drift of the grains through the gas, the non-zero 
$\Phi_p$ would not have appeared. The corresponding additional gravitational acceleration stimulates contraction
of the gas which further increases gas density and the accumulation rate of the excess dust. 
As compared to the long-wavelength as well as short-wavelength limits, dust destabilises the medium much more effectively
because of the most favourable conditions for drift, cf. equations (\ref{delta_ff}), (\ref{eq_k_high}) and 
(\ref{eq_k_J_3}).

\begin{comment}

\begin{equation}
\label{eq_k_J_1_add}
\frac{\partial^2 h^\prime}{\partial t^2} = \nabla^2 h^\prime + \tau^2 h^\prime + f \tau^2 \delta,
\end{equation}
\begin{equation}
\label{eq_k_J_2_add}
\frac{\partial \delta}{\partial t} = \tau^2 h^\prime.
\end{equation}

As soon as $k\to k_J$, the first and the second terms in RHS of equation (\ref{eq_k_J_1}) cancel each other and 
one is left with the following equation
\begin{equation}
\label{d_ttt}
\frac{\partial^3 h^\prime}{\partial t^3} \approx f \tau^4 h^\prime.
\end{equation}

%Сказать, что тут все работает именно потому, что \delta - именно превышение плотности пыли над ее же динамическим значением
%при аналогичном коллапсе газа без пыли.
%Именно за счет возмущения дрейфовой скорости (которой нет в статическом фоне, и которой нет при свободном падении вещества!)
%В этом новость найденной неустойчивости.

\end{comment}

\section{Energy of modes of gas-dust perturbations in the particular case associated with the mode coupling}
\label{sec_mode_energy}

The reduced equations describing resonance of two modes in the vicinity of the mode crossing can be
obtained from equations (\ref{eq_h_1}-\ref{eq_v}) by setting $\partial_t \approx ({\bf V\cdot \nabla})$ in the leading 
order in small $f\ll 1$ and omitting the terms $\sim f h^\prime$ in equation (\ref{eq_h_1}) and all terms $\sim f$ 
in equation (\ref{eq_delta}) after equation (\ref{eq_v}) has been used there to express $\nabla\cdot {\bf v}$. 
One finds
\begin{equation}
\label{canonical_1}
\frac{1}{c_s^2} \frac{\partial^2 h^\prime}{\partial t^2} = \nabla^2 h^\prime + \omega_{ff}^2 \frac{h^\prime}{c_s^2} + 
f \omega_{ff}^2 \delta - \frac{f}{t_s} ({\bf V}\cdot \nabla) \delta,
\end{equation}

\begin{equation}
\label{canonical_2}
\frac{\partial \delta}{\partial t} + ({\bf V}\cdot \nabla) \delta + ({\bf V}\cdot \nabla) \frac{h^\prime}{c_s^2} - 
t_s \, \omega_{ff}^2 \frac{h^\prime}{c_s^2} = 0.
\end{equation}

An additional condition\footnote{in the dimensionless units, this is $\hat V_{||} k = \tau^2$},
\begin{equation}
\label{canonical_3}
({\bf V}\cdot \nabla) = t_s\, \omega_{ff}^2,
\end{equation}
is imposed on terms that make up the coupling term in the dispersion equation, i.e. on the terms $\propto \delta$ 
in equation 
(\ref{canonical_1}) and the terms $\propto h^\prime$ in equation (\ref{canonical_2}). 
The condition (\ref{canonical_3}) reproduces to the mode coupling of HSW and SDW as it 
is a more restrictive analogue of the condition (\ref{coupl_cond}), see Section \ref{sec_coupl}.

The particular solution of equations (\ref{canonical_1}-\ref{canonical_2}) with the condition (\ref{canonical_3}) 
is taken in the following way
\begin{equation}
\label{h_1_delta}
\begin{aligned}
h^\prime = c_s^2 \tilde h^\prime \cos ( \theta - \pi/4), \\
\delta = \tilde \delta \cos \theta,
\end{aligned}
\end{equation}
where 
$$
\theta = - \omega t_* + {\bf k \cdot r_*}.
$$

The tilded quantities satisfy the following equations:
\begin{equation}
\label{eq_mod_h_1}
\omega^2 \tilde h^\prime = k^2 \tilde h^\prime - \tau^2 \tilde h^\prime - f \, \tau^2 \sqrt{2}\, \tilde \delta, 
\end{equation}
\begin{equation}
\label{eq_mod_delta}
\omega \tilde \delta = \hat {\bf V}\cdot {\bf k}\, \tilde \delta + \sqrt{2} \tau^2 \tilde h^\prime.
\end{equation}

Using the variational principle valid for modes of perturbations with the amplitude constant in time, 
which is also known as the method of \citet{whitham}, it is possible to derive the energy of gas-dust wave 
from its fundamental symmetry to translations in time. 

The averaged Lagrangian, $L(\tilde h, \tilde \delta)$, reads
\begin{equation}
\label{av_Lag}
L = (\omega^2-k^2+\tau^2) \frac{(\tilde h^\prime)^2}{2} + \sqrt{2} f \tau^2 \tilde h^\prime \tilde \delta - 
f (\omega-\hat V_{||} k) \frac{\tilde \delta^2}{2}.
\end{equation}
This Lagrangian provides equations (\ref{canonical_1}-\ref{canonical_2}) equivalent to the Euler-Lagrange equations
\begin{equation}
\begin{aligned}
\frac{\partial L}{\partial \tilde h^\prime} = 0, \\
\frac{\partial L}{\partial \tilde \delta^\prime} = 0.
\end{aligned}
\end{equation}

It can be seen that $L=0$ provided that $\tilde h_1$ and $\tilde \delta$ satisfy equations (\ref{eq_mod_h_1}-\ref{eq_mod_delta}).
Accordingly, the wave energy is as follows
\begin{equation}
\label{En_mod}
E \equiv \frac{\partial L}{\partial (\partial \theta / \partial t)} \frac{\partial \theta}{\partial t} - L = 
\omega \frac{\partial L}{\partial \omega} = \omega^2 \tilde h_1^2 - f \omega \frac{\tilde \delta^2}{2}.
\end{equation}

As $f\to 0$, eq. (\ref{En_mod}) shows that the energy of HSW and SDW is, respectively, positive and negative definite.
Indeed, the case of HSW corresponds to $\tilde \delta \to 0$, while $\omega \to (k^2-\tau^2)^{1/2}$ provided that $k>\tau$.
On the other hand, the case of SDW corresponds to $\tilde h^\prime \to 0$, while $\omega \to \hat V_{||} k$, which 
confirms that the energy SDW is negative each time the projection of the drift velocity, $\hat {\bf V}$, onto 
the wavevector of SDW is positive. 

The existence of HSW and SDW having energies of the opposite signs at the mode coupling 
allows for the standard physical explanation of instability of the gas-dust mixture in this particular case. 
Following \citet{cairns-1979} and recently \citet{zhuravlev-2019} the growing (or damping) coupled mode of gas-dust 
perturbations may be considered as the resonant coalescence of HSW and SDW, which provides an exchange with 
energy between the waves. The energy of coupled mode is conserved, while the energy flow between 
the coalescing HSW and SDW provides the growth (damping) of their amplitudes as long as the energy flows from the
negative (positive) energy wave to the positive (negative) energy wave.
As the strict condition of the mode coupling (\ref{canonical_3}) is not true, there is no simple way to construct the Lagrangian for modes of perturbations. As soon as the coupling term is complex, neutral modes do not exist for $f>0$ 
at any wavenumber. 
Moreover, according to the Lagrangian theory of perturbations developed for single-fluid dynamics, 
see \citet{friedman-1978}, the energy of growing (damping) modes must vanish. That is why, there is no straightforward 
generalisation of the standard interpretation of the resonant instability given here onto arbitrary ratio between 
$\tau$ and $\hat V_{||}$.
Note that the coupling term is real within the 'standard' concept of resonance between modes, 
see \citet{stepanyants-fabrikant-1998}. As $f>0$, the mode crossing is replaced either by
the mode coupling leading to instability or by the avoided crossing, which keeps the modes neutral (see 
\citet{zhuravlev-2019} for applications to the dynamics of gas-dust perturbations in protoplanetary disc). 
In the model of \citet{zhuravlev-2019} the coupling term becomes complex in the next order over the small stopping 
time turning one of the neutral modes taking part in the avoided crossing into a growing one. The corresponding 
mechanism of instability was referred to as 'quasi-resonant' to distinct it from the 'standard' resonant case. 
At least technically, the mechanism of the instability considered in this work also goes beyond the 'standard' case 
of the mode coupling, however, such a terminological distinction is not used in this study.

%Most likely, one should develop a generelisation of the Lagrangian approach onto non-modal perturbations.

\section{Remarks on the long-wavelength HS18 instability}
\label{sec_HS18}

The solution (\ref{sol_SH_R_3}) along with the limit (\ref{restr_SH_3_expl}) correspond to the 
following restrictions 

\begin{equation}
\label{ineq_HS18_1}
1 \gg \omega \gg k \gtrsim k \hat V_{||}
\end{equation}
along with
\begin{equation}
\label{ineq_HS18_2}
k\hat V_{||} \gg f k\hat V_{||} \gg \omega^2.
\end{equation}
Assuming $\omega_{ff} = 0$, while $t_s/t_{ev} \sim \omega$ and $c_s t_s / l_{ev} \sim k$, where $t_{ev}$ and $l_{ev}$ are, respectively, the characteristic time and scale of dynamics of the perturbed flow, equation (\ref{eq_v})
shows that perturbation of the relative velocity is determined by the acceleration of gas, i.e. 
$\nabla \cdot {\bf v} \sim \omega^2 h^\prime$. In turn, this means that the terms $\nabla^2 h^\prime$ and 
$\sim f \nabla {\bf v}$ on the RHS of equation (\ref{eq_h_1}) can be omitted as they are small compared to the main term 
on its LHS.
Thus, the acceleration of gas is determined by the aerodynamic drag emerged from the excess (deficiency) 
of dust drifting through the gas at the velocity of the bulk drift, ${\bf V}$. 
Therefore, in the long-wavelength limit, both gas and dust behave 
like pressureless fluids interacting via aerodynamic drag, which was recognised by \citetalias{squire-2018_acoustic}. 
Inequality (\ref{ineq_HS18_2}) applied to the reduced equation (\ref{eq_h_1}) indicates that, additionally, 
$h^\prime \gg \delta$, i.e. perturbation of the dust density, $\delta_p$, is
mostly generated by the compression (expansion) of gas. Moreover, inequality (\ref{ineq_HS18_2}) guaranteers that 
$\nabla\cdot {\bf v}$ can be omitted in equation (\ref{eq_delta}) as well. Thus, 
according to the reduced equation (\ref{eq_delta}), perturbation of the dust fraction is determined solely by perturbation 
of the gas density. One arrives at the following set of equations for the long-wavelength \citetalias{squire-2018_acoustic}
instability
\begin{equation}
\label{HS18_1}
\frac{\partial {\bf u}_g }{\partial t} = f \frac{{\bf V}}{t_s} \delta,
\end{equation}
\begin{equation}
\label{HS18_2}
\frac{1}{c_s^2} \frac{\partial h^\prime }{\partial t} = - \nabla \cdot {\bf u}_g,
\end{equation}
\begin{equation}
\label{HS18_3}
\frac{\partial \delta }{\partial t} = - \frac{1}{c_s^2} ({\bf V}\cdot\nabla) h^\prime.
\end{equation}

Equation (\ref{HS18_3}) shows that variations of the dust fraction emerge due to the bulk drift of the dust 
pre-compressed (pre-expanded) solidly with the gas, which is contrasted to the situation when the dust 
clumps due to the relative 
motion of gas and dust, which is excited by perturbation of the gas pressure gradient. 
The latter is the case for GI of the gas-dust mixture 
discussed in this paper, see Appendix \ref{sec_mixt_GI}, as well as for the subsonic RDI in the rotating gas-dust
mixture of protoplanetary discs, see e.g. \citet{squire_2018} and \citet{zhuravlev-2019}.

As the wave propagates through the medium with gas-dust perturbations described by equations (\ref{HS18_1}-\ref{HS18_3}), 
perturbation of the gas density is generated by the divergence of its velocity perturbation, 
which is similar to the ordinary SW. However, the feedback is different as compared to SW. 
Namely, the gas velocity perturbation is generated by the perturbation of the gas pressure gradient 
indirectly through perturbation of the dust fraction. Accordingly, oscillations of gas velocity perturbation 
are not synchronised with the force driving these oscillations. Indeed, for the ordinary SW, 
equations (\ref{HS18_1}) and (\ref{HS18_3}) are replaced by the single equation
\begin{equation}
\label{SW_eq}
\frac{\partial {\bf u}_g }{\partial t} = - \nabla h^\prime,
\end{equation}
which leads to the phase difference between ${\bf u}_g$ and $-\nabla h^\prime$ equal to $\pi/2$.
But this is not the case as the driving force is described by equations (\ref{HS18_1}) and (\ref{HS18_3}).

\begin{figure}%{h}
\begin{center}
\includegraphics[width=8cm,angle=0]{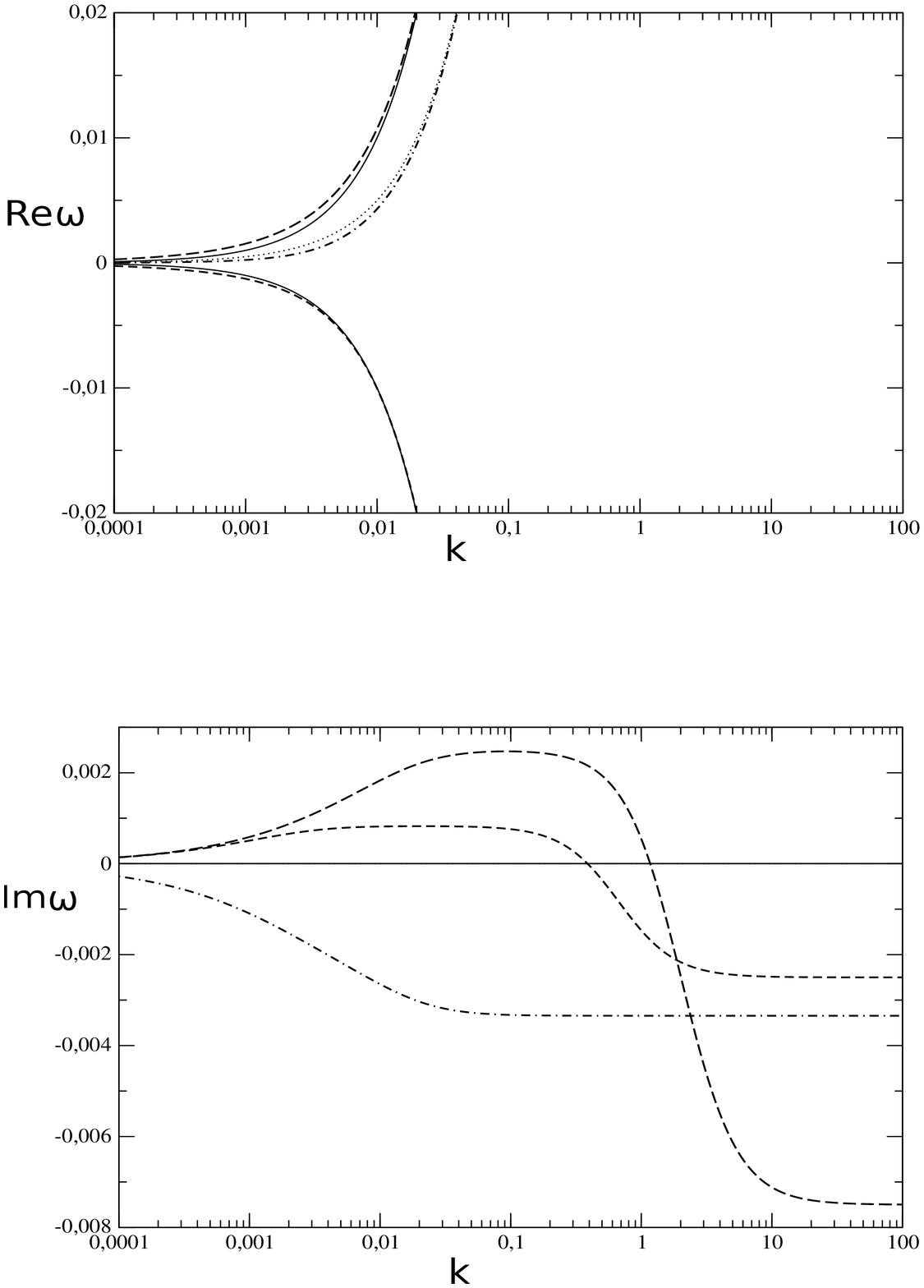}
\end{center}
\caption{The curves in top and bottom panels show, respectively, $\Re [\omega]$ and $\Im [\omega]$, 
where $\omega$ is the solution of equation ({\ref{disp}}) obtained for $\tau=0$ and $\hat V_{||}=0.5$. 
Solid and dotted lines show two SW and SDW for $f=0$, whereas short- and long-dashed lines and dot-dashed line 
show the corresponding solutions for $f=0.01$.}
\label{SH_fig}
\end{figure}

Let the gas be moving along the $x$-axis only, which at the same time be the direction of the bulk drift of the dust.
As the growing branch corresponding to 
$$
\omega = \frac{{\rm i}+\sqrt{3}}{2} f^{1/3} (\hat V_{||} k)^{2/3}
$$
is considered, it can be shown that 
$
\delta \propto \cos (\varphi -\pi/6)
$
and
$
u_g \propto \cos (\varphi + \pi/6),
$
provided that 
$
h^\prime \propto \cos \varphi,
$
where $\varphi$ is the phase of oscillations, see also \citetalias{squire-2018_acoustic}. 
The immediate cause for growth of the wave amplitude is the non-zero net work of the driving force 
acting onto the fluid elements. On the simple background given by equations (\ref{stat_2}-\ref{stat_5}) the net
work of the driving force reads
$$
A = f \frac{{\bf V}}{t_s} \oint \delta \, {\bf u}_g dt \propto \oint  \cos (\varphi -\pi/6) \cos (\varphi + \pi/6) d\varphi =\pi/2
$$
over the oscillation period of the gas element.

On the contrary, in the ordinary SW
$$
A = - \oint \nabla h^{\prime} {\bf u}_g dt \propto \oint \sin \varphi \cos \varphi = 0. 
$$

Thus, the driving force $\propto \delta$ provides the transition of the background energy of the dust drift into 
the energy of wave.

%Анализировать систему стр 43++++ записей.
%Главное, что сила, действ на газовые элементы - это не градиент давления ~ градиенту плотности газа в данном случае - 
%а пылевой фидбэк ~ \delta, который расфазирован с градиентом плотности газа (сравнивается в объяснении с устройством 
%звуковой волны), поскольку не он сам, а его производная 
%по времени определется градиентом плотности газа.
%Пылевой фидбэк генерируется на счет фонового дрейфа пыли.
%В результате, расфазировка возмущения скорости газа и пылевого фидбэка не на pi/2, как обычно в нейтралньой волне, 
% а на pi/3 (+-\pi/6 к обычной фазе в косинусе) - 
%дает вековое ускорение газовым элементам на период осцилляции - т.е. рост амплитуды волны.

%Важно, что ни при чем относительное движение пылинки-газ в возмущениях. Избыток пыли генерится фоновым дрейфом пыли 
%из областей повышенной плотности газа. 
%В этом смысле моя гравитационная неуст прямо противоположно завязана не на фоновый дрейф, а на относительное движение
%пылинки-газ в возмущениях.

For the subsonic dust drift, $\hat V <1$, the \citetalias{squire-2018_acoustic} instability ceases as $k$ approaches unity. 
An exact solution of the general equation (\ref{disp}) taken for the non-self-gravitating medium is shown in Figure
\ref{SH_fig}. It approaches (\ref{sol_SH_R_3}) in the long-wavelength limit, while each of the three branches 
seen in Figure \ref{SH_fig} become damping at its own constant rate in the limit of high $k$.
The top panel in Figure \ref{SH_fig} shows that the two curves introducing the long-wavelength \citetalias{squire-2018_acoustic}
instability approach the SW dispersion relation, while the third one, which is damping for all $k$, approaches 
the SDW dispersion relation.
In order to analytically reproduce the constant damping rates at $k\to \infty$, one should treat
the first two solutions (the dashed lines in Figure \ref{SH_fig}) as SW propagating on the dust-laden 
background with additional bulk drift of the dust, while the third solution (the dot-dashed curve in Figure \ref{SH_fig}) 
as SDW exciting the subsonic oscillations of gas.

%SW damped by the interaction of gas with dust via aerodynamic drag. 
%The third branch introduces SDW, which excites gas motion via the dust feedback. 

Let the frequency of gas-dust wave be $\omega \approx k \gg 1$. Provided $\omega \sim t_s / t_{ev}$, 
RHS of equation (\ref{eq_v}) vanishes and it yields
\begin{equation}
\label{v_with_u_g}
{\bf v} \approx - {\bf u}_g, 
\end{equation}
i.e. in the high-frequency dust-laden SW the dust velocity perturbation is negligible, ${\bf u}_p \approx 0$, 
because of the enhanced inertia of the grains. According to equation (\ref{eq_delta}), this means that
\begin{equation}
\label{delta_with_h_1}
\delta \approx -\frac{h^\prime}{c_s^2}.
\end{equation}

With help of relations (\ref{v_with_u_g}) and (\ref{delta_with_h_1}) equation (\ref{eq_h_1}) is expressed as
\begin{equation}
\label{eq_h_reduced_high_k}
\frac{\partial^2 h^\prime}{\partial t^2} = c_s^2 \nabla^2 h^\prime - 
\frac{f}{t_s} \left [ \frac{\partial h^\prime }{\partial t} + \left ( {\bf V}\cdot \nabla \right ) h^\prime \right ], 
\end{equation}
which yields the following approximate solution for $f\ll 1$:
\begin{equation}
\label{high_k_SW}
\omega \approx \pm k - \frac{{\rm i} f}{2} \left ( 1 \pm \hat V_{||} \right ).
\end{equation}
It can be checked that equation (\ref{high_k_SW}) is in good agreement with an accurate asymptotics at high $k$, 
see the dashed curves in the bottom panel of Figure \ref{SH_fig}. 

Now, let the spatially periodic perturbations of the dust density be advected by the background drift of the dust. 
Assuming that $\hat V_{||} \ll 1$ implies that the corresponding frequency of oscillations of 
aerodynamic dust feedback introduced by the last term on the RHS of equation (\ref{eq_h_1}) is small compared to the
frequency of SW with the same wavelength. Thus, the problem may be considered in the limit of $\nabla\cdot {\bf u}_g \to 0$, 
$\partial_t h^\prime/c_s^2 \to 0$, 
which implies that, according to equation (\ref{eq_v}), $\nabla\cdot {\bf v} \to 0$ and equation (\ref{eq_h_1}) 
is reduced to 
\begin{equation}
\label{eq_high_k_SDW_1}
\nabla^2 h^\prime \approx \frac{f}{t_s} ({\bf V}\cdot\nabla) \delta.
\end{equation}

In the same limit, equation (\ref{eq_delta}) reads 
\begin{equation}
\label{eq_high_k_SDW_2}
\frac{\partial \delta }{\partial t} + ({\bf V}\cdot\nabla) \delta \approx - \frac{1}{c_s^2} ({\bf V}\cdot\nabla) h^\prime.
\end{equation}
Equations (\ref{eq_high_k_SDW_1}-\ref{eq_high_k_SDW_2}) give the following solution for modes of perturbations
\begin{equation}
\label{high_k_SDW}
\omega \approx k \hat V_{||} - {\rm i} f \hat V_{||}^2,
\end{equation}
which is valid up to the leading order in the $\hat V\ll 1$.

%Поключить еще следующий порядок.
It is also possible to take into account the next order correction in $\hat V_{||}$ to estimate (\ref{high_k_SDW}).
It emerges due to the non-zero divergence of gas velocity perturbation induced by the oscillations of the dust density.
As far as the main order correction to the frequency of SDW is small, $\nabla\cdot {\bf v}$ remains to be negligible 
for the divergent gas flow as well. This can be seen from equation (\ref{eq_v}), where it is taken into account that 
$\partial_t - (V\cdot\nabla) \sim f \hat V_{||}^2$ in the dimensionless form. 
Hence, the only new term which needs to be included stands on the LHS of equation (\ref{eq_h_1}). 
Accordingly, the improved estimate of damping SDW is the following
\begin{equation}
\label{high_k_SDW_next}
\omega \approx k \hat V_{||} - {\rm i} f \frac{\hat V_{||}^2}{1-\hat V_{||}^2}.
\end{equation}
It can be checked that equation (\ref{high_k_SDW_next}) is in good agreement with an accurate asymptotics at high $k$, 
see the dot-dashed curve at the bottom panel of Figure \ref{SH_fig}.

\end{document}